\documentclass[]{aastex63}
\pdfoutput=1
\usepackage{geometry} 
\usepackage{graphicx}
\usepackage{placeins}
\usepackage{afterpage}
	
\usepackage{listings}
	
\usepackage[varg]{txfonts}
\usepackage{upgreek}
\bibpunct{(}{)}{;}{a}{}{,} 

\newcommand{\radu}{rad m$^{-2}$}

\usepackage{microtype}
\shorttitle{RMTable}
\shortauthors{Van Eck et al.}
\begin{document}
\title{RMTable2023 and PolSpectra2023: standards for reporting polarization and Faraday rotation measurements of radio sources}
\correspondingauthor{Cameron L. Van Eck}
\email{cameron.vaneck@anu.edu.au}

\author[0000-0002-7641-9946]{C.L.~Van Eck}
\affiliation{Research School of Astronomy \& Astrophysics, The Australian National University, Canberra, ACT 2611, Australia}

\author[0000-0002-3382-9558]{B.M.~Gaensler}
\affiliation{Dunlap Institute for Astronomy and Astrophysics, University of Toronto, 50 St. George Street, Toronto, ON M5S 3H4, Canada}

\author[0000-0002-6952-9688]{S.~Hutschenreuter}
\affiliation{Department of Astrophysics/IMAPP, Radboud University, P.O. Box 9010,6500 GL Nijmegen, The Netherlands}
\affiliation{Max Planck Institute for Astrophysics, Karl-Schwarzschildstr. 1, 85741 Garching, Germany}
\affiliation{Ludwig-Maximilians-Universität München, Geschwister-Scholl-Platz 1, 80539 Munich, Germany}

\author[0000-0002-4090-8000]{J.~Livingston}
\affiliation{Research School of Astronomy \& Astrophysics, The Australian National University, Canberra, ACT 2611, Australia}

\author[0000-0003-0742-2006]{Y.K.~Ma}
\affiliation{Research School of Astronomy \& Astrophysics, The Australian National University, Canberra, ACT 2611, Australia}
\affiliation{Max-Planck-Institut für Radioastronomie, Auf dem Hügel 69, D-53121 Bonn, Germany}

\author[0000-0002-3369-1085]{C.J.~Riseley}
\affiliation{Dipartimento di Fisica e Astronomia, Universit\`a degli Studi di Bologna, via P. Gobetti 93/2, 40129 Bologna, Italy}
\affiliation{INAF -- Istituto di Radioastronomia, via P. Gobetti 101, 40129 Bologna, Italy}
\affiliation{CSIRO, Space \& Astronomy, PO Box 1130, Bentley, WA 6102, Australia}

\author[0000-0001-9472-041X]{A.J.M.~Thomson}
\affiliation{CSIRO, Space \& Astronomy, PO Box 1130, Bentley, WA 6102, Australia}

\author[ 0000-0002-5447-6878 ]{B.~Adebahr}
\affiliation{Faculty of Physics and Astronomy, Ruhr University Bochum, Astronomical Institute, Universitätsstr 150, D-44801 Bochum, Germany}
\affiliation{ASTRON, the Netherlands Institute for Radio Astronomy, Oude Hoogeveensedijk 4, 7991 PD Dwingeloo, The Netherlands}

\author[0000-0003-2030-3394]{A.~Basu}
\affiliation{Thüringer Landessternwarte, Sternwarte 5, D-07778 Tautenburg, Germany}
\affiliation{Max-Planck-Institut für Radioastronomie, Auf dem Hügel 69, D-53121 Bonn, Germany}

\author[0000-0002-1858-277X]{M.~Birkinshaw}
\affiliation{H.H. Wills Physics Laboratory, University of Bristol, Bristol BS8 1TL, UK}

\author[0000-0001-5246-1624]{T. A.~Enßlin}
\affiliation{Max Planck Institute for Astrophysics, Karl-Schwarzschildstr. 1, 85741 Garching, Germany}

\author[0000-0002-2155-6054]{G.~Heald}
\affiliation{CSIRO, Space \& Astronomy, PO Box 1130, Bentley, WA 6102, Australia}

\author[0000-0001-8906-7866]{S.A.~Mao}
\affiliation{Max-Planck-Institut für Radioastronomie, Auf dem Hügel 69, D-53121 Bonn, Germany}

\author[0000-0003-2730-957X]{N.M.~McClure-Griffiths}
\affiliation{Research School of Astronomy \& Astrophysics, The Australian National University, Canberra, ACT 2611, Australia}

\begin{abstract}
Faraday rotation measures (RMs) have been used for many studies of cosmic magnetism, and in most cases having more RMs is beneficial for those studies. This has lead to development of RM surveys that have produced large catalogs, as well as meta-catalogs collecting RMs from many different publications. However, it has been difficult to take full advantage of all these RMs as the individual catalogs have been published in many different places, and in many different formats. In addition, the polarization spectra used to determine these RMs are rarely published, limiting the ability to re-analyze data as new methods or additional observations become available.\\
We propose a standard convention for RM catalogs, RMTable2023, and a standard for source-integrated polarized spectra of radio sources, PolSpectra2023. These standards are intended to maximize the value and utility of these data for researchers and to make them easier to access. To demonstrate the use of the RMTable2023 standard, we have produced a consolidated catalog of 55 819 RMs collected from 42 published catalogs.

\end{abstract}

\section{Introduction}
Magnetic fields are a fundamental component of nature, and one which plays an important role in many aspects of astrophysics. Magnetic fields have been found to influence cosmic ray propagation through galaxies \citep{Farrar16}, cloud collapse in star formation \citep{VanLoo12}, pressure balance in the interstellar medium \citep{Boulares90}, and more. Magnetic fields in astrophysical environments are often probed using the polarization of light, as there are several physical processes where information on the magnetic field is imprinted on the polarization state of light emitted from or passing through the magnetized environment. At radio wavelengths, synchrotron emission and Faraday rotation are two such processes that are measured to provide information on the magnetic fields in astrophysical systems.

Radio spectropolarimetry is required for measurements of Faraday rotation, due to the frequency dependent nature of the Faraday effect in magnetized plasma. While early observations measured only a few frequencies \citep[e.g.,][]{Cooper1962, Rudnick1983}, improvements in radio telescopes and computational resources have enabled observations with hundreds to thousands of frequency channels covering increasingly wide fractional bandwidths. A key quantity that is often derived from radio spectropolarimetry is the Faraday rotation measure (RM), which gives the amount of Faraday rotation between the source of the polarized emission and the observer. By measuring RMs for many lines of sight to compact sources that pass through a region of interest, it becomes possible to model the line-of-sight magnetic field in the region; this has been applied to the Galactic disk \citep{Brown2003} and halo \citep{Mao2012Halo}, high velocity clouds \citep{Betti2019}, molecular clouds \citep{Tahani2018}, nearby galaxies \citep{Heald09,Mao2012LMC}, galaxy clusters \citep{Bonafede2010}, intervening galaxies \citep{Farnes2017}, and more.

RM studies have often focused on polarized sources that are compact (unresolved, or marginally resolved) to reduce complications involved in disentangling the effects of source structure on the polarization properties. The majority of these sources are extragalactic (typically radio galaxies and active galactic nuclei), but pulsars are often also significantly polarized and can have measured RMs \citep[e.g.,][]{Sobey2021}. While there have been many catalogs of RM values published over the years, there has been very little consistency in the contents of these catalogs; by definition an RM catalog need only contain the positions of the polarized radio sources and their measured RMs (in some cases, even the uncertainty in RM is not reported). Many other properties are often reported, such as source brightness, polarized fraction, polarization angle, and more, but it is essentially left to the interests and whims of the catalog authors as to which are included in any given catalog. This potentially limits the value of the catalog for future users, depending on the set of properties those users need for their analyses.

Since there have been many surveys and independent projects measuring RMs over the past decades, there has been a long history of RM compilations or meta-catalogs collecting many RM catalogs together \citep{Tabara1980, Broten1988, Klein2003, Xu2014}. These compilations provide significant value by saving users the work of finding the various individual catalogs in the literature, with the cost that the focus is generally only on RM while additional polarization information (e.g., polarized fraction or polarization angle), uncertainty estimates, and metadata about the specifics of the individual observations are often discarded. Limiting the amount of ancillary information present, both in the original publication of the catalogs and in the catalog compilations, reduces the legacy value of the catalogs by limiting the kinds of analyses for which those data can be used in the future. Therefore there is an incentive for the astronomical community to push catalog authors to include a wider set of relevant parameters in their catalogs where reasonably possible.

As radio observations have advanced, RM catalogs have evolved from being typically a handful of sources observed with only a few frequency channels, to containing up to hundreds or thousands of sources each observed with hundreds of channels. This increase in catalog size is in large part due to a trend towards larger survey projects, as well as advancements in telescope and computing technology. As these surveys become larger, with more telescope time and more human resources being devoted into their execution and processing, it becomes increasingly important to maximize the value of the data products produced by these surveys, by ensuring that they are useful for as broad a range of possible studies as reasonably possible. The next generation of radio polarization surveys includes the POlarization Sky Survey of the Universe's Magnetism \citep[POSSUM, ][]{Gaensler2010}, which expects to produce RMs for approximately $10^6$ sources, as well as polarization products from the Very Large Array Sky Survey \citep[VLASS,][]{Lacy2020, Mao2014}, the LOFAR Two-metre Sky Survey \citep[LoTSS, ][]{Shimwell2017}, the MeerKAT International GHz Tiered Extragalactic Exploration \citep[MIGHTEE, ][]{Jarvis2016}, Spectra and Polarization in Cutouts of Extragalactic sources from the Rapid ASKAP Continuum Survey (SPICE-RACS), and surveys conducted using the Apertif system \citep{vanCappellen2022} on the Westerbork Synthesis Radio Telescope (WSRT). These in turn will lead to future polarization surveys with the Square Kilometre Array \citep{Heald2020}.

Since polarized spectra contain more information than simply the RM, there have also been efforts to publish these spectra \citep[e.g.,][]{Klein2003, Farnes2014, OSullivan2017}. These offer significant scientific value by enabling later studies using the data for such purposes as reproducibility, alternative analysis by other authors, or applying newly developed analysis methods to existing data. However as the size and complexity of radio datasets have increased, this practise has become the exception rather than the norm, in part due to the complexity of storing and representing these data in an efficient way.

There are significant benefits to the scientific community by investing effort into ensuring that published data follows best practices \citep{Chen2022}. While some of these best practices are very widespread (e.g., including uncertainties on all measured quantities), they remain not universal (e.g., \citealt{Kim2016} reported RMs with no uncertainties); other best practices are much more uncommon, such as including important observation metadata  (e.g., observation dates, which we found in only approximately one fourth of published RM catalogs).

It is clear that the radio polarization community would find significant value in having standardized formats for presenting and sharing their results. In this paper, we propose two data standards for representing polarization results in a consistent way, one for RM catalogs and one for full Stokes spectra of radio sources. We also present examples of software written to work with data following these standards. In Section~\ref{sec:standard} we present the standard for RM catalogs, along with a Python package to enable easy use of this standard, and in Section~\ref{sec:catalog} describe a catalog, following this standard, of previously published RMs. Section~\ref{sec:polspectra} describes a standard for storing polarized spectra of radio sources and its intended usage, along with a Python package to interact with data following this standard. Finally in Section~\ref{sec:summary} we summarize the key aspects of these standards and discuss how the community can best make use of them.

\subsection{Faraday rotation terminology}
To avoid possible ambiguity in the standards defined in the following sections, it is useful to carefully define the concepts and terminology relevant to Faraday rotation studies. Such ambiguities have been seen before in the literature, as there are (at least) four related quantities that have been described using two terms: rotation measure and Faraday depth.

For a given line of sight, the degree of Faraday rotation experienced between a distance $d$ and the observer is given\footnote{This formulation ignores cosmological redshift effects. A revised equation for Faraday rotation that accounts for redshift can be found in, e.g., \citealt{Vallee1975}.}
by 
\begin{equation}\label{eq:FRstrength}
\phi(d) = 0.812\; \mathrm{rad \; m}^{-2} \int_0^d \frac{n_e}{\mathrm{cm}^{-3}}\, \frac{B_\parallel}{\upmu\mathrm{G}} \, \frac{dr}{\mathrm{pc}}
\end{equation}
where $n_e$ is the free electron density, $B_\parallel$ is the component of the magnetic field parallel to the line of sight, and $dr$ is an element along the line of sight (taken as positive towards the source). The parallel magnetic field is taken as positive when directed towards the observer \citep{Ferriere2021}. The quantity $\phi(d)$ is generally called the Faraday depth, and is physically defined for any location in space independent of any source of polarized emission. The emission from a polarized source at some single distance is said to have an RM equal to the Faraday depth at that distance, although in many cases this cannot be straightforwardly assigned.

Faraday rotation causes the polarization angle, $\chi$, to be changed by an amount $\Delta\chi = \lambda^2 \, \phi(d)$, where $\lambda$ is the emission's observed wavelength. When a line of sight has (or is assumed to have) a single source of polarized emission at a single distance, the RM can be observationally derived through several different methods, such as by measuring or fitting the relationship between polarization angle and wavelength-squared \citep[e.g., ][]{Brown2003}. Other methods, such as RM-Synthesis \citep{Burn66, Brentjens2005}, characterize the polarized emission in the space of all possible Faraday depths (the Faraday dispersion function, FDF), and define the RM as some characteristic Faraday depth associated with that function (e.g., the value of Faraday depth at which polarized intensity is found to be maximum).

Assigning an RM to a source is generally predicated on the assumption that all the polarized emission of the source has a single value of Faraday depth.
This requires that either all the emission is emitted at the same distance, or the emitting volume all has the same value of Faraday depth (i.e., no Faraday rotation internal to the emitting volume). This is referred to as the (ideal) Faraday-simple case. Cases where a line of sight has polarized emission distributed over a range in Faraday depth, either through Faraday rotation in the emitting volume or by having multiple emission regions at different distances within the same resolution element, are called Faraday-complex. The sensitivity of an observation to Faraday complexity is strongly dependent on the frequency coverage of the data and width of the emission in Faraday depth, so sources can appear to be effectively Faraday-simple in some observations and as Faraday-complex (to differing degrees) in other observations. In general a Faraday-complex source cannot be assigned a single value for RM, but in certain limiting cases one can choose to assign RM values to such sources. For example, if a line of sight has multiple polarized emission features that are clearly distinct in Faraday depth, an RM value can be assigned for each feature \citep[e.g.,][]{Schnitzeler2019}. Also, if the Faraday depth distribution of a polarized emission feature has a clear centroid, that value may be reported as the RM of the source \citep[e.g., ][]{Anderson2016}.

\section{RM catalog standard: RMTable2023}\label{sec:standard}
A standard for rotation measure catalogs offers several advantages to the scientific community. The largest benefit may be that authors of new catalogs can use the standard as a guide for what measurable quantities may be of scientific interest beyond those authors' immediate interests, prompting them to add more information than they otherwise would. The second benefit is to make it much more straightforward to combine multiple catalogs together for an analysis, something which is currently either difficult or not worthwhile given the very heterogenous nature of existing published catalogs. A third benefit is that it can encourage more standardization in how RMs are derived (and the reporting thereof; e.g. the use of ionospheric correction), increasing the compatibility of different data sets.

The standard presented here was developed as the combination of two separate efforts: planning of the catalog design for the POSSUM and VLASS projects, and a proposal to assemble a new meta-catalog of published RMs. The POSSUM survey is expected to produce approximately one million RMs, and the VLASS data may produce as many as 250 000 RMs; the pipeline and data product planning for these surveys required defining the catalog contents. In parallel, we began working on a new collection, or meta-catalog, of published RMs be compiled for use in all-sky statistical studies of Faraday rotation \citep{Hutschenreuter2022}, and faced questions regarding how such a meta-catalog should be structured. The obvious overlap between these two efforts, combined with the difficulties faced in assembling the meta-catalog, provided clear motivation for developing a general catalog format suitable for most future RM catalogs.

The catalog standard proposed below, which we call `RMTable2023',\footnote{We have included the year of publication in the standard name to explicitly distinguish it from possible future revisions.} 
consists of three parts: a statement of scope defining the kind of data products for which this standard is suitable, a set of column definitions for the suggested contents of an RM catalog, and suggested data formats for efficiently storing and manipulating such a catalog. This standard is not intended to be restrictive or proscriptive as there may be aspects that need to be modified to suit specific projects, although the greater the deviation from the standard the less applicable the benefits described above become. The most common example, and one that the standard was designed to support, is the addition of extra columns specific to a particular project, such as observation information, source cross-identifications, redshifts, or any other parameters that authors may wish to include.


\subsection{Scope}
The intended scope for RMTable2023 is this: it can be used for tabulating the polarization and Faraday rotation measurements for all discrete radio sources with one or more characteristic RM values. The specification of discrete sources is to exclude diffuse Galactic emission, which varies strongly with position on the sky, but to include sources that are resolved (or partially resolved) but can still be assigned an RM. It is intended to include multiple spatially-resolved components within a larger source (e.g., individual lobes or hotspots within a radio galaxy), so long as they can be interpreted as identifiable discrete regions with a well-defined RM value for each, but it is intended to exclude sources with smooth spatial variations in RM. Resolved objects may also be included, in cases where authors assign a single RM value to such objects. The exact boundary between these conditions is at the discretion of individual catalog authors, but the proposed test for suitability is: if it is better described by an RM map than a table of discrete components, then it is not suited for the RMTable2023 standard.

The restriction to sources with `one or more characteristic RM values' is made in recognition of the capability of modern algorithms, such as RM synthesis \citep{Brentjens2005} and QU-fitting \citep[e.g., ][]{OSullivan2012}, to identify and characterize Faraday complexity within sources. The most general case, where the polarized emission is distributed over Faraday depth with some arbitrary distribution that isn't reducible to a small number of characteristic Faraday depth values \citep[e.g., many of the FDFs shown in][]{Basu2019}, is outside the scope of the standard. However, most analyses model or fit the distribution as a combination of a small number of components each with a well defined Faraday depth centroid which is generally taken as the RM for that component. Given the development of surveys that present more than one component for a source \citep[e.g.,][]{OSullivan2017,Schnitzeler2019}, it is necessary for any RM catalog standard to support this. While it is possible to represent this by allowing columns such as RM to contain multiple values, this greatly increases the complexity of storing and interacting with the catalogs. The simpler solution, which we expand on and endorse below, is to have each RM become a separate row in the catalog, resulting in sources with multiple polarized components appearing in multiple catalog entries (with common properties, such as position, Stokes $I$, observation parameters, etc. repeated for each component).

Similarly, in cases where a source has multiple reported RMs at different times, the proposed solution is to have each measurement (and other information) as separate rows in a catalog table. Each row should contain a single RM determination: where such determinations use observations from multiple epochs this should be reflected in the time-related columns of the catalog, but where separate RM determinations have been done (at different times, with different data, or possibly with different methods) these should be recorded as separate rows in the catalog table. In some catalogs (e.g., a consolidated catalog spanning a wide time-range of observations), it may be expected for individual sources to have multiple entries, which may or may not constitute independent measurements depending on the observations used for each.

\subsection{Column definitions}
The RMTable2023 standard contains a set of proposed standard columns to guide the development of future RM catalogs. The list of columns is given in Table~\ref{tab:RMTable}, and specific information about each column is given below. Not all of these columns will be appropriate for all catalogs, and individual catalogs will almost always need to have additional columns with survey- or catalog-specific information, so these columns serve as a starting point for authors to consider while designing their catalogs and pipelines. The columns defined in this standard, and particularly their internal names, should be treated as reserved names for future catalogs to avoid confusion.

All columns have standardized internal names, which are used in the data representation to assign machine-readable names to columns. Relevant columns also have assigned standard units, to avoid problems with propagating units through appropriate metadata, and to ease comparisons between different catalogs. A few columns (those defining the source sky coordinates) are labelled as essential, meaning that they must be present for a catalog to be sensibly interpreted in terms of the standard. All other columns are assigned a default value for missing data, which is usually a floating-point NaN or blank string as appropriate. 
To encourage standardization across catalogs, certain columns (particularly those defining methods used) come with lists of values that have been used in existing catalogs or are anticipated to be used in near-future catalogs. The currently defined lists are given in Appendix~\ref{app:standards}, but an updated, living version will be kept in the same location as the Python module described in Section~\ref{sec:RMTable}. These values are suggested for improving consistency between catalogs, but catalog authors are free to define new values as needed.

For columns specified as being floating point numbers, it is not specified whether these should be 32- or 64-bit floats. The exception to this is the coordinate columns, which are specified as 64-bit (double precision) floating point in order to accommodate RMs derived from Very Long Baseline Interferometry observations.
All uncertainties/errors are expected to be 1$\sigma$ standard deviations. All string columns should take care not to contain special characters such as tabs or newlines, to avoid causing problems with text storage formats.
Where columns are expected to refer to published papers, the suggested format is to use `bibcodes' created by the SAO/NASA Astrophysics Data System (ADS), with the second preferred option being digital object identifiers (DOIs); these are suggested because they provide an identifier that is guaranteed to be unique and can easily lead to the publication using either the ADS or an internet search engine, with the ADS bibcodes producing typically fewer false hits when searching and being partially human-readable. For cases where the papers are not yet published a short descriptive name can be used as a temporary measure but should be replaced after publication.
All columns are defined the same way for both RM synthesis techniques and QU-fitting techniques, although the methods of deriving some quantities may be significantly different.

\subsubsection{Right Ascension, Declination, Galactic longitude and latitude}
The sources' coordinates, in both equatorial and Galactic coordinate systems, in decimal degrees (for ease of coordinate-based selection). Both sets of coordinates are considered essential in order to make it easy to select sources based on location. For equatorial coordinates, the International Celestial Reference System (ICRS) is used as it is the currently adopted standard of the International Astronomical Union.

\subsubsection{Position error}
The 1-dimensional uncertainty in the position, in decimal degrees. Commonly, position errors are reported on both position coordinates, but this neglects the possible covariance between the uncertainties (or equivalently, the orientation of the uncertainty ellipse, if it is not aligned with one of the coordinate axes). This makes it impossible to accurately transform the uncertainties between coordinate systems, or to assess the significance of source position differences if not aligned with the coordinate axes. There is also an ambiguity as to whether errors in RA or Galactic longitude are great-circle distances or coordinate distances. To avoid all of these issues, the standard uses a single value to describe a circular position uncertainty (although this comes with the disadvantage of being less accurate where the position error ellipse is significantly elongated). In cases where the position error ellipse is significantly non-circular, we suggest using the semi-major axis (i.e. the larger error) for this column in order to be conservative with the uncertainty values. The standard does not prevent catalog authors from including either the full position uncertainty ellipse or the errors in the each coordinate, but we suggest that in such cases authors also include this column to maximize ease of combination with other RMTable2023 catalogs.

\subsubsection{RM and error}
The RM and corresponding error, in \radu. For Faraday-simple polarized sources, this is simply the measured Faraday depth of the polarized emission. As described above, in the case of Faraday-thick features this is the centroid in Faraday depth or a similar quantity. For partially spatially-resolved sources this is the average over the source, or whatever other characteristic value the authors decide applies to the source. If multiple polarization components are present, each is recorded as a separate row in the table with its own RM, error, and other polarization properties. The catalog value should always represent the astrophysical RM of the full line of sight to the source as measured; RMs with any component or foreground subtraction (e.g., removing the Milky Way contribution) or redshift correction should not be reported. If a catalog author wishes to report a corrected RM, this should be as a separate column; including the 'raw' RM values allows catalog users to perform other types of analysis (e.g., Galactic magnetism studies) or apply a correction of their choice (e.g., using a newer foreground model). The exception to this is the subtraction of ionospheric Faraday rotation (discussed below) as such subtraction is often done early in data processing (to allow time-dependent corrections), preventing any calculation of a non-corrected RM.

\subsubsection{Width in Faraday depth and error}
The width of the polarized feature in the FDF, and corresponding error, in \radu. This can be either measured (from RM synthesis) or inferred from a model (QU-fitting). This quantity is zero when an explicitly Faraday-thin model is fit or assumed, and can take other values when either a Faraday thick model is fit \citep[e.g.,][]{Ma2019} or the dispersion in RM-synthesis clean components is measured \citep[e.g.,][]{Livingston2021}. It must be noted that there are many (incompatible) measures of width in Faraday depth: the $\Delta$RM of a uniform `Burn slab' model \citep{Burn66} is not equivalent to the Gaussian $\sigma_{\mathrm{RM}}$ of an external Faraday dispersion model \citep{Sokoloff98} or to a $\sigma^2_\phi$ derived from RM-synthesis clean components. This is further complicated by cases where two width parameters are included in the same component fit \citep[e.g., ][]{OSullivan2017}; in such cases authors may need to determine a method of combining these parameters into a single value. In general this quantity will not be comparable between models and between different catalogs, but may be useful to compare sources within the same catalog. Authors should ensure that the width parameter is explicitly defined, and catalog users should refer back to the original publications to find these definitions.

\subsubsection{Faraday complexity flag and metric}
A single parameter for whether the source shows Faraday complexity, which for the purposes of this standard is anything other than being compatible with a single-component Faraday-thin model. An additional value, for sources where the complexity has not been assessed or reported, is also needed. Thus three possible values are needed, and a single character is used with the possible values of `Y',`N`, or `U' for `Yes', `No', and `Unknown', respectively. For sources with two or more RM components or a component with a significant non-zero width in Faraday depth, this flag should automatically become `Y' (for each row, where multiple components are present). The motivation for including this column is to provide users with an easy way to identify the Faraday simple or complex sources present in the catalog without needing to check for additional components or significant Faraday width.

The metric or test used to assess Faraday complexity is also recorded as a short string. This string could be a reference to a paper describing the metric, or a short description (preferrably 80 characters or less) of the method. A list of currently used or suggested values appears in Appendix~\ref{app:standards} and will continue to be updated in the online documentation as new values come into use, in order to encourage authors to use the same values when using the same methods.

\subsubsection{RM determination method}
A short string describing the method used to determine the RM from the polarized spectra. This can be useful when using a consolidated catalog to assess differences between measurements of the same source; with QU-fitting this column is used to specify the exact model used. A list of currently used or suggested values appears in Appendix~\ref{app:standards} and will continue to be updated in the online documentation as new values come into use, in order to encourage authors to use the same values when using the same methods.

\subsubsection{Ionospheric correction method}
A string describing the method used to correct for ionospheric Faraday rotation. This could take the form of the software title, a paper reference, or a short algorithm name. A list of currently used or suggested values appears in Appendix~\ref{app:standards} and will continue to be updated in the online documentation as new values come into use, in order to encourage authors to use the same values when using the same methods. If no correction is applied, the value should be `None'; if it is not known if a correction was applied the default value is `Unknown'. This column can be useful for catalog users to assess the possible presence of systematic RM offsets in the catalog due to ionospheric effects.

\subsubsection{Number of RM components}
An integer giving the number of measured RM components identified in the source. This is intended to indicate when the same source will have additional rows in the table (containing the information on the other components). The default value is `1'.

\subsubsection{Stokes $I, Q, U, V$ and errors}
The four Stokes parameters and their corresponding errors, at a given reference frequency. Stokes $I$ may have a different reference frequency, as it may be derived from multi-frequency synthesis imaging or some other algorithm that is different than how the other Stokes parameters are derived. The Stokes parameters may be either intensities (Jy/beam) or flux densities (solid-angle integrated intensities, Jy). Brightness temperatures are strongly discouraged, as such values are rarely appropriate for discrete sources. For Stokes $Q$ and $U$, these are the values derived from the Faraday rotation model (whether that model is a QU-fitting model, an RM-Clean \citep{Heald09} component model, a fit to the peak in RM-synthesis, or something analogous to these), at the corresponding reference frequency and for only this polarized component, and not the actual channel values.

\subsubsection{Stokes $I$ spectral index and error}
The flux density spectral index ($\alpha$) in total intensity, following the $I \propto \nu^{+\alpha}$ convention (such that most synchrotron sources will have negative values), and corresponding error. This can be the in-band spectral index or that derived with an additional band. Higher order (curvature) terms are not included in this standard, but where authors choose to include them we suggest that the mathematical definition of those terms should be explicitly described. Providing the spectral index can be very useful for users looking to classify sources or select certain source populations.

\subsubsection{Reference frequency for Stokes $I$}
The frequency, in Hz, for which the Stokes $I$ intensity or flux density applies, as well as for the spectral index if a spectral curvature model was fit. This value is necessary for users to understand and compare the Stokes I values to observations made at other frequencies.

\subsubsection{Polarized intensity and error}
The polarized intensity (in either intensity or flux density units) of the polarized component, at the polarization reference frequency. If polarization bias correction is used, this is the corrected polarized intensity. In sources where multiple polarized components have been identified, this is the polarized intensity of only the component with the corresponding RM. In the Faraday-complex case polarized intensity is not well defined, as there are multiple ways it could be defined depending on the method being used. In RM-synthesis methods this could be the amplitude of the peak of the FDF or some form of integrated measurement over the Faraday depth range of the component. In QU-fitting methods, this is most commonly either the polarized intensity after all depolarizing effects are removed (which is a direct fitting parameter in most models) or the polarized intensity extrapolated to zero wavelength (which is the same as the former in most models). In polarization angle linear fitting, this may be the average of polarized intensity across all the channels. Catalog authors should describe how they defined the polarized intensity when publishing their catalogs; catalog users should carefully check this definitions before using the values, especially if comparing values across multiple catalogs.

\subsubsection{Polarization bias correction method}
A string containing the method used to correct the polarized intensity for bias \citep[e.g., ][]{Wardle1974, Simmons1985}. If no correction was applied, this should be `None'; if it is not known if a correction was applied, this should be `Unknown'. A list of currently used or suggested values appears in Appendix~\ref{app:standards} and will continue to be updated in the online documentation as new values come into use, in order to encourage authors to use the same values when using the same methods. Reporting this value in a catalog is important to allow users to assess the possible effects of polarization bias on the polarized intensity values in the catalog.

\subsubsection{Stokes extraction method}
A string describing the method used to extract the source spectra, for example whether they were intensities derived from peak-pixel values, aperture-integrated flux densities, or intensities or flux densities derived from point-source or Gaussian fitting. If the method is not known, the default value is `Unknown'. A list of currently used or suggested values appears in Appendix~\ref{app:standards} and will continue to be updated in the online documentation as new values come into use, in order to encourage authors to use the same values when using the same methods. Including this column in a catalog can be helpful for users when trying to evaluate possible causes of differences between different measurements/catalogs or to interpret results from partially resolved sources.

\subsubsection{Integration aperture}
The linear size of the integration aperture over which the spectra have been integrated or averaged, in decimal degrees. If only peak/single pixel values are extracted from the images, this should be zero. If a circular or square aperture is used, the diameter or side length should be given. If a Gaussian-fit or similar process was used, then the FWHM of the fitted area would be appropriate. This information can be useful when trying to analyze a partially resolved source or to reproduce a catalog result.

\subsubsection{Fractional (linear) polarization and error}
The fractional polarization of the polarized component, and corresponding error. This parameter suffers from the same ambiguities and multiple possible definitions as polarized intensity; see the description of the polarized intensity column for a description of the complications. Values for this column should be fractional, and not percentages; nominally values should be less than 1, but this is not enforced because of the complications that can potentially be introduced by interferometer response and other effects. These columns can be very useful for studies of depolarization in sources (e.g., by comparing fractional polarization across different frequencies) as well as evaluating possible effects of polarization leakage.

\subsubsection{Electric vector polarization angle and error}
The electric vector position angle (EVPA, or simply polarization angle) and corresponding error, in degrees and at the polarization reference frequency, following the IAU standard convention \citep{IAU-polangle}: increasing east from north, with zero degrees being towards the north celestial pole. We note that cosmology data sometimes use a different convention \citep{deSeregoAlighieri2017}, west-from-north, so care should be used in correcting the convention when using such data. Where Stokes $Q$ and $U$ follow the IAU convention, the EVPA is defined as $\frac{1}{2}\tan^{-1}\frac{U}{Q}$. Polarization angles should not be reported relative to the Galactic coordinate frame; the angles should also be explicitly the EVPA and not the plane-of-sky magnetic field. Polarization angle is only defined over a 180$^\circ$ span; we have chosen [0\degr,180\degr) rather than the sometimes-used ($-90$\degr,90\degr].\footnote{Note that the 180\degr\ span occurs after the division by two in the definition above. When calculating the polarization angle, it is important to use an arctan function that can return all possible angles (e.g., the `atan2' function in many programming languages, and not the `atan' function which only spans half the angle range).}

\subsubsection{Reference frequency for polarization}
The frequency, in Hz, at which the relevant polarization properties are applicable. This applies to the polarized intensity, EVPA, and the Stokes $Q,U$, and $V$ values. For RM synthesis techniques, this frequency typically corresponds to the parameter $\lambda^2_0$ \citep{Brentjens2005}; for QU-fitting there is no equivalent value and authors can choose a suitable value (we suggest to make this equal to the Stokes I reference frequency) or leave the corresponding columns blank.

\subsubsection{De-rotated EVPA and error}
The EVPA with the effects of Faraday rotation removed (i.e. the polarization angle at the location of emission), and associated error, in degrees. This is sometimes called the `zero wavelength' or `intrinsic' polarization angle. This value can be useful in cases when users are interested in the magnetic field orientation (for synchrotron-emitting sources).

\subsubsection{Beam major axis, minor axis, and position angle}
The three parameters describing the shape of the synthesized beam at the reference frequency, as a Gaussian: the major axis, minor axis, and position angle, all in degrees. The major and minor axes are the FWHM of the Gaussian beam model, along the major and minor axes respectively, and the position angle is the angle of the major axis measured east from north similarly to the polarization angle, and is similarly defined in the range [0\degr\,180\degr). This information can be very useful when comparing results with different studies that have differing resolution, if a source may be (partially) resolved in one or more of the observations.

\subsubsection{Reference frequency for beam shape}
The reference frequency for the beam shape parameters, in Hz, if the beam size follows a typical $1/\nu$ frequency dependence. If the individual channels have been convolved to a common size, this frequency should be set to zero to indicate that the beam has no frequency dependence. If the frequency dependence of the beam is not known, this should be left blank (NaN). This value can be useful as an indicator of whether frequency-dependent resolution could be producing any effects on the results, as well as allowing users to determine beam effects at different frequencies.

\subsubsection{Lowest and highest frequencies}
The center frequencies of the lowest and highest frequency channels used in determining the RM, in Hz. Channels that have been flagged out or otherwise not used should not be used to determine these values. These values can be useful for users who wish to assess the expected Faraday depth resolution and sensitivity to Faraday-complex signals that are broad in Faraday depth for a given observation.

\subsubsection{Typical channel width}
The bandwidth of the channels used to determine the RM, in Hz. If channels were averaged before being used to compute the RM, the width of the averaged channels should be used. If channels of different widths have been used together, this should be the most common channel bandwidth (the mode). Channel bandwidth is the key determining factor in bandwidth depolarization, so this value is helpful for assessing whether an observation may be biased against large RMs.

\subsubsection{Number of channels}
The number of frequency channels used to determine the RM. Channels that have been flagged out or otherwise were not used should not be included in this count. Since integers cannot use NaN values to represent missing data, any negative number can be used to represent missing values. This column can be useful in conjunction with the channel width to estimate total bandwidth, or to assess the ability of the data to constrain QU-fitting models with many parameters.

\subsubsection{Full width at half max of the rotation measure spread function}
The FWHM of the rotation measure spread function (RMSF) calculated during RM-synthesis, in rad m$^{-2}$. If RM-synthesis is not used, this column can be ignored or set to the theoretical RMSF FWHM (as defined in \citealt{Brentjens2005}) given the frequency coverage of the data. This column can provide a convenient metric for users to assess the Faraday depth resolution and sensitivity to Faraday complexity of an observation.

\subsubsection{Typical per-channel noise in Stokes $Q,U$}
An estimate for the noise in Stokes $Q$ and $U$ for a typical channel, in the same units as the Stokes parameters. The exact method of determining this is left to the catalog authors, but a mean or median of the channel noise values would be reasonable. This value can be useful as a metric of data quality, and for users planning deeper observations of interesting sources.

\subsubsection{Name of Telescope(s)}
A string containing the names or acronyms of all telescopes from which data were used, as a comma separated list. A list of currently used or suggested values appears in Appendix~\ref{app:standards} and will continue to be updated in the online documentation as new values come into use, in order to encourage authors to use the same values. The default value, if the data origin is not known, is `Unknown'.

\subsubsection{Integration time}
The integration time is the amount of time the telescope spent observing the source, in seconds, for a typical channel. If multiple observations at the same frequency were combined, then this should be the sum of the individual observation integration times, but if the observations were for different frequencies then the mean, median, or mode of the integration time for the individual channels should be used. This value can be useful for users concerned about possible time-averaging effects.

\subsubsection{Median epoch and interval of observation}
The median epoch is the midpoint of time between the first and last observations used to determine the RM. If a single observation was used, it should be the time at which the observation was half-complete. This time is stored as the modified Julian date (MJD, JD-2,400,000.5). The interval of observation is the span of time between the beginning of the first observation and the end of the final observation used to determine the RM, in days. If only a single observation was used, this is the difference between the start and end times of that observation. These columns allow RMs to be used in analysis of the evolution of RM over time.

\subsubsection{Instrumental leakage estimate}
An estimate of the degree of instrumental leakage present in Stokes $Q$ and $U$, expressed as a fraction of Stokes $I$. If a leakage correction has been applied, this should be an estimate of the residual leakage after correction. This information can be useful to assess the significance of a detection (i.e., the risk of a reported RM being due to instrumental leakage rather than the astrophysical source), as well as the possible degree of systematic error introduced by leakage (which is distinct from the random error which is usually reported for quantities like Stokes $Q$ and $U$ or polarized intensity).

\subsubsection{Distance from beam center}
The angular distance of the source from the primary beam center in the observations, in degrees. If multiple observations or a phased array feed are used, this should be the distance from the nearest beam center. Since this quantity can be an indication of the possible severity of off-axis leakage effects, its purpose is to allow users to qualitatively estimate the relative effect of leakage on different sources (especially if the instrumental leakage estimate column is not supplied).

\subsubsection{Name of catalog}
A string containing a unique name for the catalog. The first preference for this is the paper in which the catalog was published, following the usual preference for the ADS bibcode or DOI. If the paper is not yet published, a short descriptive name can be used as a temporary substitute.

\subsubsection{Data references}
A string containing references to the sources of data used in determining the RM, following the preferences for references described above. If the paper reporting the catalog also reports on the observations used, then this should be the same paper. If data from multiple papers are used, a comma-separated list should be used. This column can be critically important for determining if RMs published in different catalogs are not independent (i.e., calculated from the same observations).

\subsubsection{Source ID in catalog}
A string containing the name of the source used in the catalog, if any. It is left completely to the authors' discretion whether this is a name unique to their catalog, a source name/id from another catalog (e.g., a 3C number), or something else.

\subsubsection{Source Classification}
A string containing the source classification, specifically what kind of physical object the source is. If multiple classifications have been given, each should be separated by a comma. A list of currently used or suggested values appears in Appendix~\ref{app:standards} and will continue to be updated in the online documentation as new values come into use, in order to encourage authors to use the same values.

\subsubsection{Notes}
A string containing any short notes the authors have made about individual sources.

\startlongtable
\begin{deluxetable*}{lllccll}
\tablecaption{Column definitions for the RMTable2023 standard.\label{tab:RMTable}}
\tablehead{\colhead{Column Name} & \colhead{Internal name} &  \colhead{Data format} & \colhead{Unit} & \colhead{Limits} & \colhead{Default/Missing}}
\startdata
\multicolumn{6}{c}{\bf Position columns:}\\
Right Ascension [ICRS] & ra & double & deg & [0,360) & Essential   \\
Declination [ICRS] & dec & double & deg & [-90,90] & Essential  \\
Galactic Longitude & l & double & deg & [0,360) & Essential  \\
Galactic Latitude & b & double & deg & [-90,90] & Essential  \\
Position uncertainty & pos\_err & float & deg & [0,$\infty$) & NaN  \\
\multicolumn{6}{c}{\bf RM columns:}\\
Rotation measure & rm & float & rad m$^{-2}$ &(-$\infty$,$\infty$) & NaN\\
Error in RM & rm\_err & float & rad m$^{-2}$ &[0,$\infty$) & NaN\\
Width in Faraday depth & rm\_width & float & rad m$^{-2}$ & [0,$\infty$)& NaN\\
Error in width & rm\_width\_err & float & rad m$^{-2}$& [0,$\infty$)& NaN\\
Faraday complexity flag & complex\_flag & string & -- & `Y'/'N'/`U' & `U' \\
Faraday complexity metric$^a$ & complex\_test & string & -- & -- & `'\\
RM determination method$^a$ & rm\_method & string & -- & -- & `Unknown' \\
Ionospheric correction method$^a$ & ionosphere & string & -- & -- & `Unknown'\\
Number of RM components & Ncomp & integer & -- & [1,$\infty$) & 1\\
\multicolumn{6}{c}{\bf Polarization properties:}\\
Stokes I$^b$ & stokesI & float & Jy or Jy/beam & [0,$\infty$) & NaN\\
Error in Stokes I$^b$ & stokesI\_err & float & Jy or Jy/beam & [0,$\infty$) & NaN\\
Stokes I spectral index & spectral\_index & float & -- & (-$\infty$,$\infty$) & NaN\\
Error in spectral index & spectral\_index\_err & float & --  & [0,$\infty$) & NaN\\
Reference frequency for Stokes I & reffreq\_I & float & Hz & (0,$\infty$) & NaN\\
Polarized intensity & polint & float & Jy or Jy/beam & [0,$\infty$) & NaN\\
Error in Pol.Int. & polint\_err & float & Jy or Jy/beam & [0,$\infty$) & NaN\\
Polarization bias correction method$^a$ & pol\_bias & string & -- & -- & `Unknown'\\
Stokes extraction method$^a$ & flux\_type & string & -- & -- & `Unknown'\\
Integration aperture & aperture & float & deg & [0,$\infty$) & NaN\\
Fractional (linear) polarization & fracpol & float & -- & [0,$\infty$) & NaN\\
Error in fractional polarization & fracpol\_err & float & -- & [0,$\infty$) & NaN\\
Electric vector polarization angle & polangle & float & deg & [0,180) & NaN\\
Error in EVPA & polangle\_err & float & deg & [0,$\infty$) & NaN\\
Reference frequency for polarization & reffreq\_pol & float & Hz & (0,$\infty$) & NaN\\
Stokes $Q^b$ & stokesQ & float & Jy or Jy/beam & (-$\infty$,$\infty$) & NaN\\
Error in Stokes $Q^b$ & stokesQ\_err & float & Jy or Jy/beam & [0,$\infty$) & NaN\\
Stokes $U^b$ & stokesU & float & Jy or Jy/beam & (-$\infty$,$\infty$) & NaN\\
Error in Stokes $U^b$ & stokesU\_err & float & Jy or Jy/beam & [0,$\infty$) & NaN\\
De-rotated EVPA & derot\_polangle & float & deg & [0,180) & NaN\\
Error in De-rotated EVPA & derot\_polangle\_err & float & deg & [0,$\infty$) & NaN\\
Stokes $V^b$ & stokesV & float & Jy or Jy/beam & (-$\infty$,$\infty$) & NaN\\
Error in Stokes $V^b$ & stokesV\_err & float & Jy or Jy/beam & [0,$\infty$) & NaN\\
\multicolumn{6}{c}{\bf Observation properties:}\\
Beam major axis & beam\_maj & float & deg & [0,$\infty$) & NaN\\
Beam minor axis & beam\_min & float & deg & [0,$\infty$) & NaN\\
Beam position angle & beam\_pa & float & deg & [0,180) & NaN\\
Reference frequency for beam & reffreq\_beam & float & Hz & [0,$\infty$) & NaN\\
Lowest frequency & minfreq & float & Hz & (0,$\infty$) & NaN\\
Highest frequency & maxfreq & float & Hz & (0,$\infty$) & NaN\\
Typical channel width & channelwidth & float & Hz & (0,$\infty$) & NaN\\
Number of channels & Nchan & integer & -- & (0,$\infty$) & Any negative integer$^c$\\
Full-width at half maximum of the RMSF & rmsf\_fwhm & float & rad m$^{-2}$ &[0,$\infty$) & NaN\\
Typical per-channel noise in $Q,U^b$ & noise\_chan & float & Jy or Jy/beam & [0,$\infty$) & NaN \\
Name of Telescope(s)$^a$ & telescope & string & -- & -- & `Unknown'\\
Integration time & int\_time & float & s & [0,$\infty$) & NaN \\
Median epoch of observation & epoch & float & days & (-$\infty$,$\infty$) & NaN\\
Interval of observation & interval & float & days & [0,$\infty$) & NaN \\
Instrumental leakage estimate & leakage & float & -- & [0,$\infty$) & NaN\\
Distance from beam center & beamdist & float & deg &[0,$\infty$) & NaN\\
\multicolumn{6}{c}{\bf Miscellaneous:}\\
Name of catalog & catalog & string & -- & -- & Essential\\
Data references & dataref & string & -- & -- & `'\\
Source ID in catalog & cat\_id & string & -- & -- & `'\\
Source classification$^a$ & type & string & -- & -- & `'\\
Notes & notes & string & -- & -- & `'\\
\enddata
\ \tablecomments{Columns marked as essential are required to have a value and cannot be blank.\\
See text for additional notes on some columns.\\
$^a$: These columns have a list of currently used or suggested values to encourage standardization.\\
$^b$: All Stokes parameters can be either flux densities or intensities.\\
$^c$: Since NaN is not generally defined for integers, a negative integer should be used to represent missing data. The default behaviour in Python is to replace NaNs with -2147483648, so this value is generally used in RMTables generated by Python.}
 \end{deluxetable*}

\subsection{Suggested file formats}
The RMTable2023 standard is not prescriptive about the file formats that can be used to store RM catalogs, in order to remain flexible as common practices in astronomy evolve. However, a few comments can be made about appropriate choices for file formats. Both text (ASCII, Unicode, or other) and binary formats can be used for catalogs, without a clear advantage for either. Binary formats, such as Flexible Image Transport System (FITS) binary tables \citep{Cotton1995}, are more efficient for storing the numerical data, but rely more on specific software tools to access and manipulate the file contents. Text files are less efficient at storing numerical data but can shorten strings (if a non-fixed width format is used), while being very simple to access through a wide variety of tools. 

Fixed-width text formats, which have been popular in astronomy for many years, are discouraged for RMTables as they generally use fixed decimal expansions that do not cope well with data that span many orders of magnitude. While individual RM catalogs will generally have similar values for RM, error in RM, Stokes parameters, etc. for most sources, different catalogs can span many orders of magnitude in those quantities, making it difficult or inefficient for a fixed-width format to capture all values without truncation losses. Of the commonly-used non-fixed width formats, comma separated values (CSV) is not suited for RMTables, as some columns in the standard may contain comma separated lists which would interfere with the file format. Alternative column separators, such as tab-separated values (TSV), should be effective in general.

The RMTable2023 standard is fully compatible with the VOTable format \citep{VOTable}, which allows both binary or text encoding and provides a mechanism for including additional metadata if desired.

The software package described in Sec.~\ref{sec:RMTable} can write out an RMTable in three formats: FITS binary table, which is compatible with the {\em Astropy} package \citep{Astropy} and {\em TOPCAT} software \citep{TOPCAT}; TSV text format, which is compatible with many ASCII table parsers; and VOTable format (using the `TABLEDATA' serialization, which stores values as text within an eXtensible Markup Language (XML) framework), which is compatible with {\em TOPCAT} and other Virtual Observatory compatible tools.

\subsection{RMTable package}\label{sec:RMTable}
To support the use of the RMTable2023 standard, we have developed a Python 3 package, also called RMTable, for the creation and manipulation of RM catalogs. This package is intended to make it easier for authors to build their catalogs following the standard, and easier for users to read such tables into Python for interaction. The core of the package is built around an RMTables class, which in turn is built on top of an {\em Astropy} Tables instance with additional code to supply the standard column names and default values and to check that all supplied values are consistent with the expected limits and conventions. Functions are included for the following tasks:
\begin{itemize}
\item read and write catalogs into a FITS binary table format, TSV text format, and VOTable format
\item convert the table into a {\em Numpy} \citep{Numpy} structured array
\item convert the table into a {\em pandas} \citep{pandas} dataframe
\item convert  {\em Numpy} tables into RMTables, assigning default values to any missing columns
\item verify that data in the table conform to the standard, including limits and suggested standard string values
\item add additional columns beyond the standard
\item calculate missing coordinates, if only one of Equatorial or Galactic are supplied
\item merge tables together (including reconciling differences in columns present in each table)
\item extract specified rows and/or columns from the table.
\end{itemize}

Appendix~\ref{app:RMTexample} shows examples of how this package can be used to generate and manipulate an RMTable object. A living/updated version of the code for the package is hosted on Github\footnote{https://github.com/CIRADA-Tools/RMTable} and is available through the Python Package Index (PyPI). The GitHub repository will also maintain information on any updates to the standard. A permanent version of the code at time of publication will be archived with the journal and on the arXiv.

\subsection{Suggestions for RM catalog authors}
Publishing a catalog following the RMTable2023 standard should not be a significant burden for most authors; in most cases it could be achieved with only a few hours of additional effort.\footnote{In the process of assembling the consolidated catalog described in Sec.~\ref{sec:catalog}, previously published catalogs were converted to RMTable2023 format with a typical time investment of less than one hour per catalog.} The majority of the columns defined in the standard are optional and can be omitted or left blank without creating problems, and in many cases some of the columns will not applicable to some catalogs, although every column that is included increases the value of the catalog. The key minimum elements that must be adhered to follow the RMTable2023 standard are twofold: first, the standard columns that {\em are} included must use the naming convention and units of the standard (to avoid users being unable to combine catalogs, or combining catalogs with inconsistent units); second, any columns added that are outside the RMTable2023 standard must not have a name conflict with any of the defined standard columns (e.g., a column labelled ``b'' would conflict with the Galactic Latitude column in RMTable2023). As long as those two conditions are satisfied, catalog authors have the freedom to choose how much effort they invest into including more of the standard columns.

For future catalog authors who are in the early stages of processing observations to produce an RM catalog, we suggest that it can be very valuable to review the list of RMTable2023 columns to see which values can be easily extracted by data processing pipelines. It can often be the case that values can be extracted early in the initial processing (e.g., observation dates, channel parameters) and propagated through to the final catalog much more easily than trying to cross-reference individual sources in the final catalog back to their original data. It can also be the case that authors can identify columns that they would not otherwise have computed or reported but that can be added to their processing pipelines with minimal effort. In general, we expect authors to use their best judgement to decide which standard columns can be included without undue time investment.

\section{Consolidated RM catalog}\label{sec:catalog}
To demonstrate the utility of the RMTable2023 standard, and to provide a valuable resource to the community, we have assembled a new consolidated catalog of published RMs drawn from 42 publications. This catalog was assembled from published catalogs available in machine-readable format that were within the scope of the standard as defined in Sec.~\ref{sec:standard}.

Converting existing catalogs into the RMTable2023 standard required parsing the machine-readable tables and assigning columns from these tables to corresponding columns in the standard, converting units as required. Columns that were not present in the tables but were given in the text of the catalog papers, such as the frequencies and telescope(s) used, were added manually. As a result, adding new catalogs required some effort and it was not possible to convert and incorporate all known RM catalogs. We prioritized papers based on three factors. First, we set aside catalogs of pulsar RMs, even though they are within the scope, as pulsars have their own consolidated catalog\footnote{\citet{Manchester2005}, https://www.atnf.csiro.au/research/pulsar/psrcat/} with RMs. Fast radio burst (FRB) RMs have also not yet been incorporated for the same reason\footnote{\citet{Petroff2016}, https://www.frbcat.org/}, but could be in the future. Second, we prioritized larger catalogs over smaller ones; since the amount of work required to incorporate a catalog did not scale with catalog size, this was the most effective way to get the largest catalog for a fixed amount of time invested. Third, we prioritized newer catalogs over older ones, on the principle that newer data tend to be higher quality and more relevant for modern studies.

\begin{figure*}[p]
    \centering
    \includegraphics[width=\textheight,height=0.9\linewidth,keepaspectratio,angle=90]{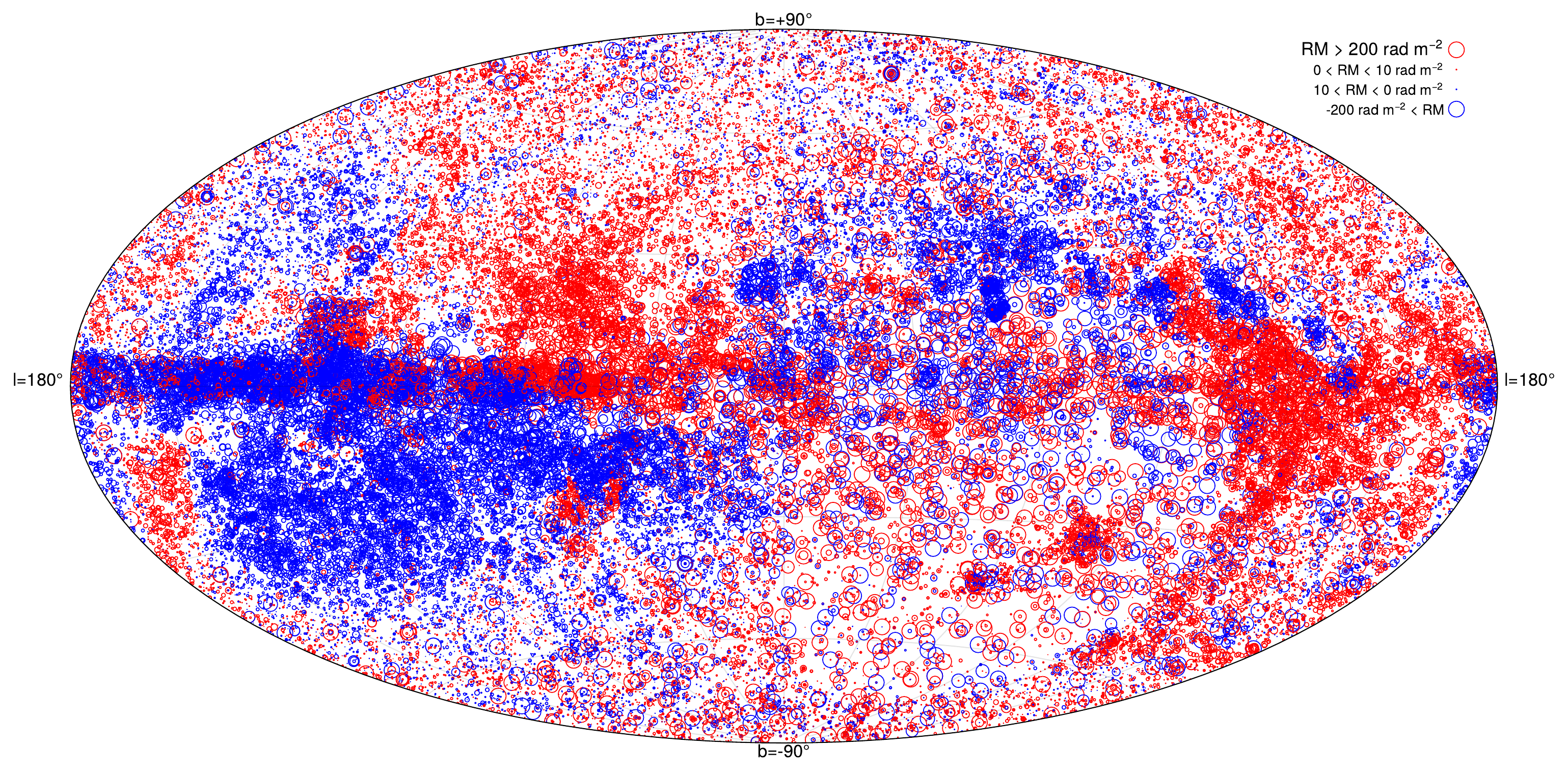}
    \caption{All-sky distribution of RMs in Galactic coordinates, plotting each source as a circle with diameter proportional to magnitude of RM (capped at 200 \radu) centred on the position of the source; positive RMs are shown in red and negative in blue.}
    \label{fig:circle}
\end{figure*}

\begin{figure*}[p]
    \centering
    \includegraphics[width=\linewidth,height=0.9\textheight,keepaspectratio]{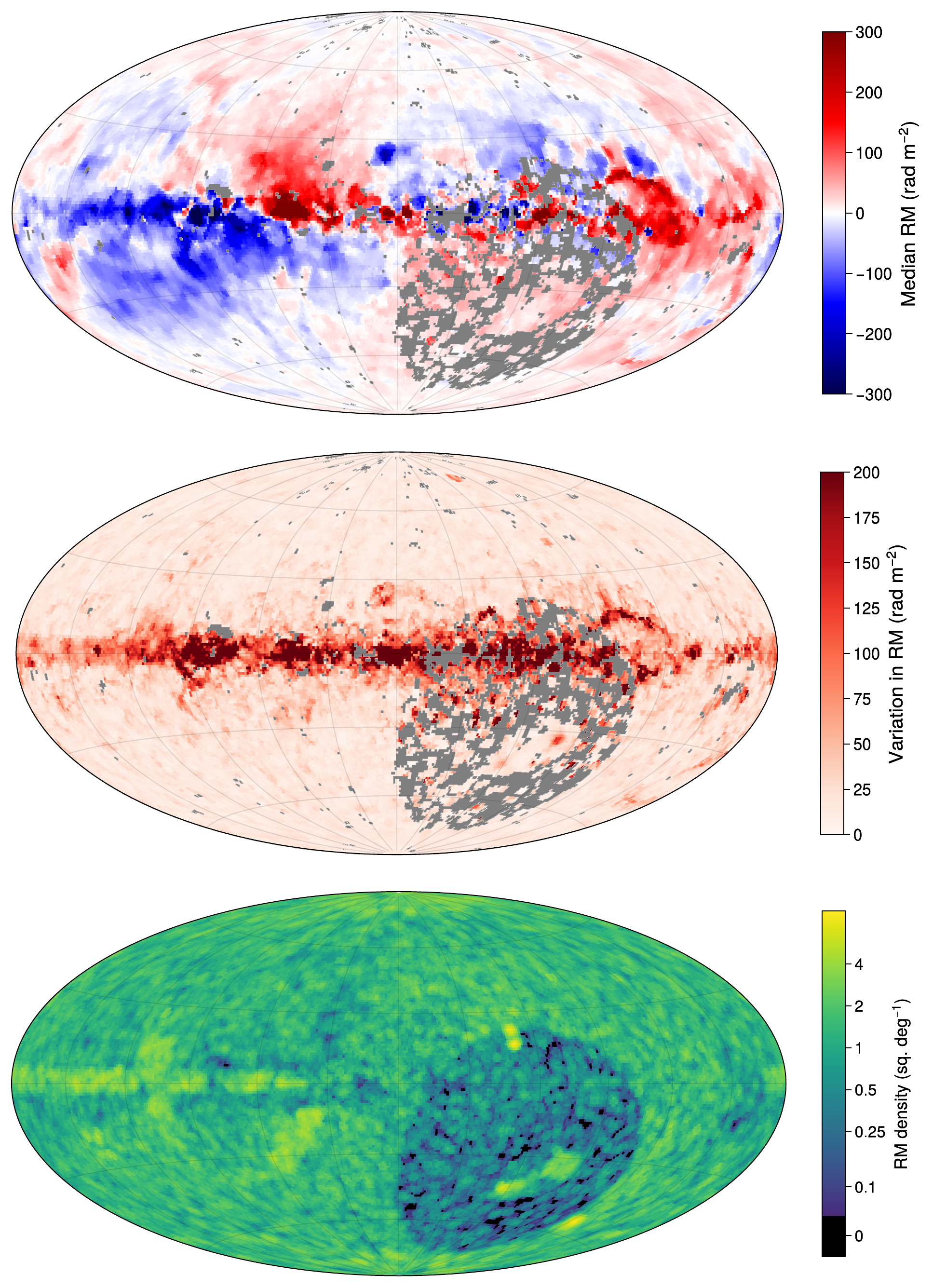}
    \caption{All-sky distributions of local median RM, variation of RM, and RM density, in a Hammer projection in Galactic coordinates (Galactic center at the center). The median and variation are computed as the median and median absolute deviation from the median (MADFM), normalized to be equivalent to the standard deviation, within a 2\degr\ radius of each point; regions where there were fewer than 5 sources within 2\degr\ were left blank (marked in grey). The source density was likewise computed locally over a 2\degr\ radius, regions with no sources within 2\degr\ are black.}
    \label{fig:allsky}
\end{figure*}

\begin{figure}[p]
    \centering
    \includegraphics[width=\linewidth,height=0.9\textheight,keepaspectratio]{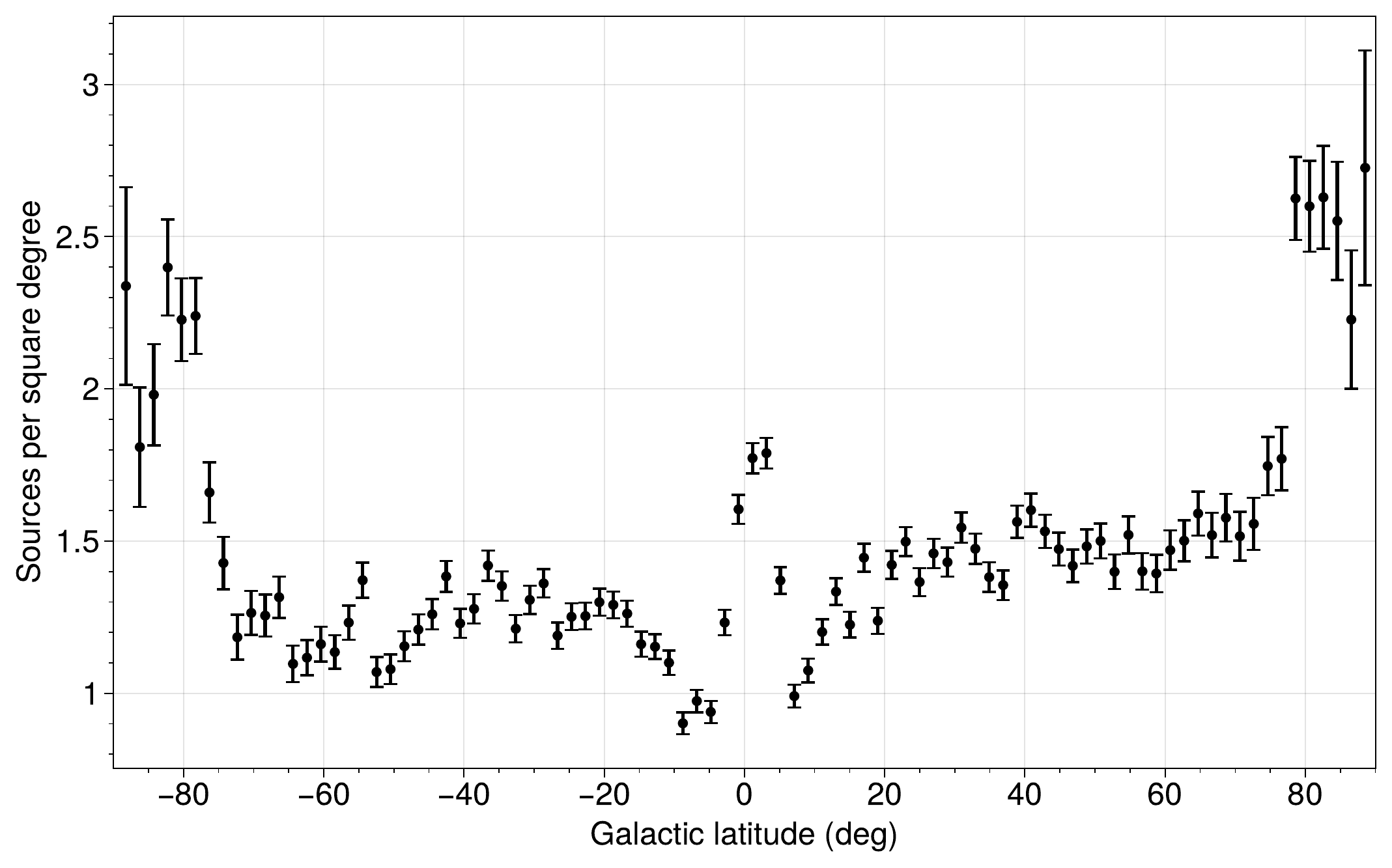}
    \caption{RM density as a function of Galactic latitude, with 2 degree bins. The enhanced density at the Galactic poles is due to targeted surveys \citep{Mao2010}. There is an under-density of sources above and below the Galactic plane in the range 5\degr $< |b| <$ 10\degr, and an over-density of sources at the Galactic plane ($|b| < $ 2\degr).}
    \label{fig:latitude}
\end{figure}

The list of catalogs that have been incorporated into the current version of the consolidated catalog is given in Table~\ref{tab:papers}. Appendix \ref{app:papers} contains notes on any complicating factors, changes, or usage notes that may apply to the individual catalogs that were incorporated. Together these catalogs contain 55~819 RMs. Figure~\ref{fig:circle} shows the position and RMs of all the sources in the catalog. Figure~\ref{fig:allsky} shows the distribution of RM and RM variation across the sky, as well as the local sky areal density of RMs in the catalog. We note that this is not the best determination of the Galactic RM contribution; more sophisticated modelling such as by \citet{Oppermann2012} and \citet{Hutschenreuter2022} is generally more robust than directly sampling the local RM population. The low density of RMs at low declinations, below the declination limit of the \citet{Taylor09} catalog ($\delta$ = $-40$\degr), produces a clear gap in the lower right portion of each panel, although targeted surveys of the Small and Large Magellanic Clouds and Centaurus A fill in part of that region. There is also a significant deficit of sources near the Galactic plane; Figure~\ref{fig:latitude} shows the Galactic latitude distribution of sources and the density of sources within 5\degr $< |b| <$ 10\degr\ is significantly lower than the mid-plane density and the density farther from the plane. It is not clear to what degree this is a sampling effect (a combination of Galactic plane surveys targeting $|b| \lesssim 5\degr$ and other surveys avoiding the confusion of the plane) versus a physical effect (greater depolarization by differential Faraday rotation in the plane reducing detected source counts). 

\begin{figure}[p]
    \centering
    \includegraphics[width=\linewidth,height=0.9\textheight,keepaspectratio]{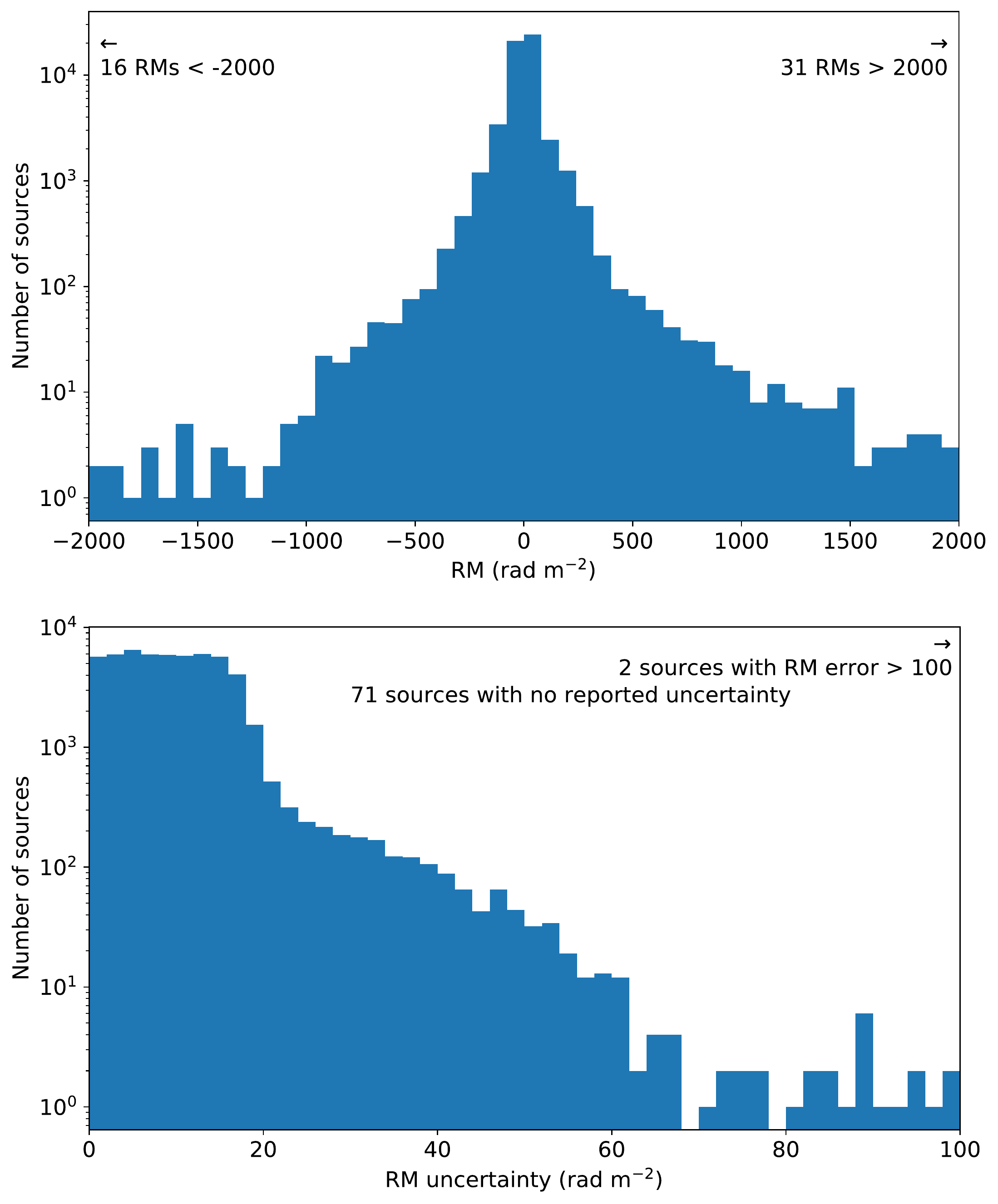}
    \caption{Histograms showing the catalog's distribution of RMs (top panel) and measurement uncertainty in RM (bottom panel).}
    \label{fig:histograms}
\end{figure}

Figure~\ref{fig:histograms} shows the distribution of RM and uncertainty in RM for the entire catalog. The unusual distribution of the RM uncertainties (specifically, the uniform distribution of uncertainties below 18 \radu) has been verified to come from the reported uncertainties and not from a transcription error in compiling the catalog; it is not clear what factors have led to this distribution. 

\begin{figure}[p]
    \centering
    \includegraphics[width=\linewidth,height=0.9\textheight,keepaspectratio]{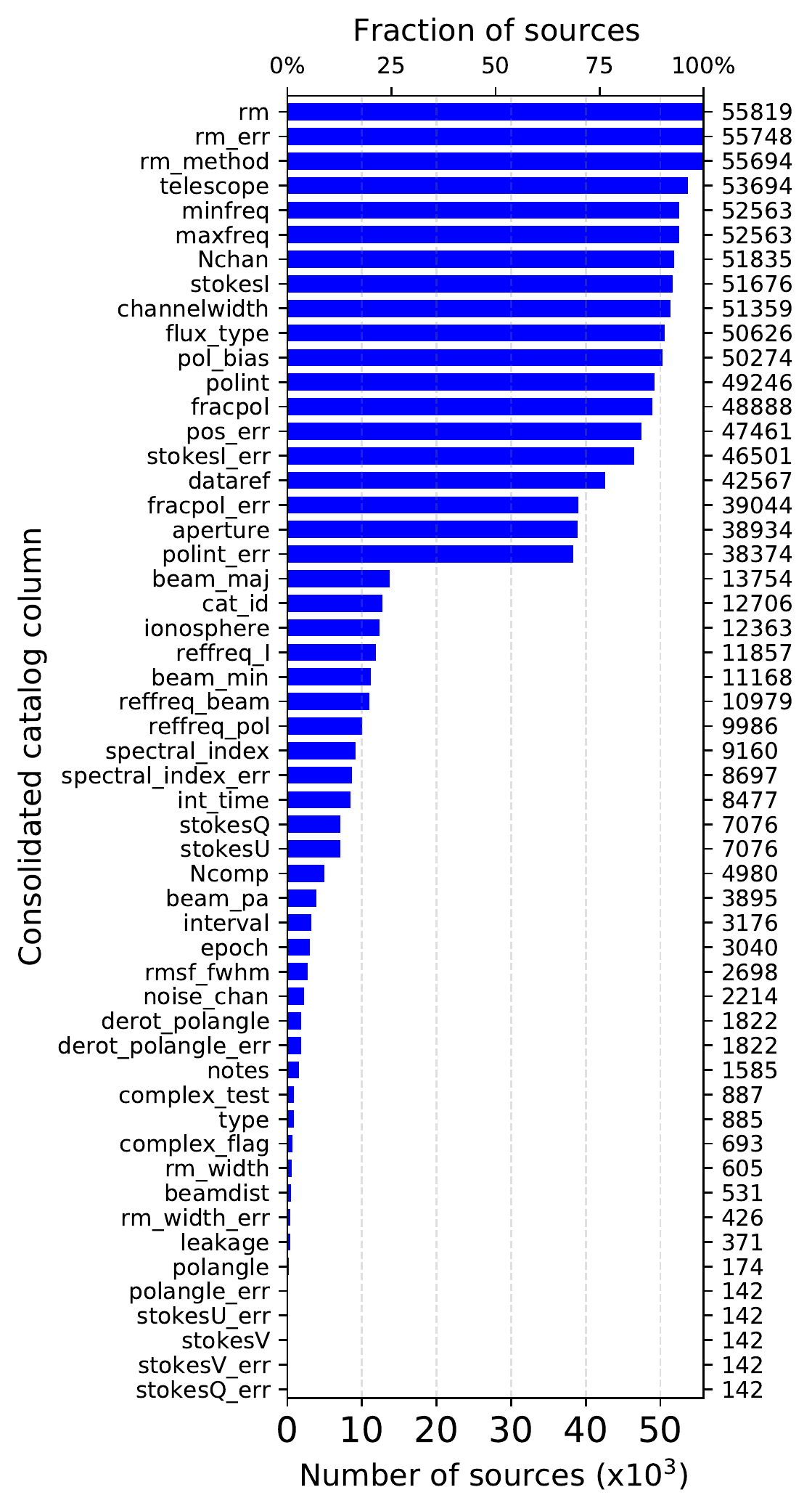}
    \caption{The number of entries in the consolidated catalog with (non-default/blank) values for each column; the exact number is given on the right axis. The full catalog contains 55 819 sources.}
    \label{fig:completeness}
\end{figure}

Figure~\ref{fig:completeness} shows, for each of the non-essential columns, how many sources possess non-default values. Nearly all sources in the catalog have uncertainties in RM, and most have Stokes $I$, polarized intensity, and fractional polarization measurements with uncertainties. The other quantities defined in the standard have not been commonly reported in previous catalogs.

\begin{deluxetable*}{rrl}
\tablecaption{List of catalogs included in the consolidated catalog, ordered by catalog size.}
\tablehead{\colhead{Catalog reference} & \colhead{\# of sources} & \colhead{Catalog ID}}
\startdata
\citet{Taylor09} & 37 543 & 2009ApJ...702.1230T\\
\citet{Schnitzeler2019} & 6 934 & 2019MNRAS.485.1293S\\
\citet{VanEck2021} & 2 234 & 2021ApJS..253...48V\\
\citet{Betti2019} & 1 105 & 2019ApJ...871..215B\\
\citet{Farnes2014} & 907 & 2014ApJS..212...15F\\
\citet{Mao2010} & 813 & 2010ApJ...714.1170M\\
\citet{Tabara1980}$^a$ & 704 & 1980A\&AS...39..379T\\
\citet{Broten1988}$^{a,b}$ & 672 & 1988Ap\&SS.141..303B\\
\citet{Simard-Normandin1981} & 555 & 1981ApJS...45...97S\\
\citet{Riseley2020} & 516 & 2020PASA...37...29R\\
\citet{Brown2003} & 380 & 2003ApJS..145..213B\\
\citet{Mao2012LMC} & 305 & 2012ApJ...759...25M\\
\citet{Mao2012Halo} & 302 & 2012ApJ...755...21M\\
\citet{Feain2009} & 281 & 2009ApJ...707..114F\\
\citet{VanEck11} & 194 & 2011ApJ...728...97V\\
\citet{Ma2020} & 194 & 2020MNRAS.497.3097M\\
\citet{OSullivan2017}$^c$ & 174 & 2017MNRAS.469.4034O\\
\citet{Kaczmarek2017} & 167 & 2017MNRAS.467.1776K\\
\citet{Anderson2015} & 160 & 2015ApJ...815...49A\\
\citet{Brown07} & 148 & 2007ApJ...663..258B\\
\citet{Klein2003} & 143 & 2003A\&A...406..579K\\
\citet{Heald09}$^c$ & 133 & 2009A\&A...503..409H\\
\citet{Shanahan2019} & 127 & 2019ApJ...887L...7S\\
\citet{Clarke2001} & 125 & 2001ApJ...547L.111C\\
\citet{Minter1996}$^b$ & 98 & 1996ApJ...458..194M\\
\citet{VanEck2018a} & 92 & 2018A\&A...613A..58V\\
\citet{Law2011} & 90 & 2011ApJ...728...57L\\
\citet{Riseley2018} & 81 & 2018PASA...35...43R\\
\citet{Livingston2022} & 80 & 2022MNRAS.510..260L\\
\citet{Mao2008} & 70 & 2008ApJ...688.1029M\\
\citet{Roy2005} & 67 & 2005MNRAS.360.1305R\\
\citet{Livingston2021} & 62 & 2021MNRAS.502.3814L\\
\citet{Oren1995}$^b$ & 61 & 1995ApJ...445..624O\\
\citet{Clegg1992}$^b$ & 56 & 1992ApJ...386..143C\\
\citet{Kim2016} & 49 & 2016ApJ...829..133K\\
\citet{Battye2011} & 45 & 2011MNRAS.413..132B\\
\citet{Ma2019} & 35 & 2019MNRAS.487.3432M\\
\citet{Rossetti2008} & 32 & 2008A\&A...487..865R\\
\citet{Costa2018} & 27 & 2018ApJ...865...65C\\
\citet{Vernstrom2018} & 22 & 2018MNRAS.475.1736V\\
\citet{Gaensler2001} & 21 & 2001ApJ...549..959G\\
\citet{Costa2016} & 15 & 2016ApJ...821...92C\\
\hline
{\bf Total: } & 55 819 &
\enddata
\tablecomments{
a. These papers are older collections of previously published RMs, which we have incorporated directly to avoid the difficulty of finding the many original catalogs (many of which do not exist in machine-readable form).\\
b. The coordinate and RM data for these catalogs were taken from a machine-readable consolidated catalog compiled by Jo-Anne Brown.\footnote[4]{http://www.ras.ucalgary.ca/~jocat/RMData/}\\
c. These catalogs presented multiple RMs/polarized components per sources. We have split each component into its own row in the consolidated catalog.}
\label{tab:papers}
\end{deluxetable*}

\subsection{Catalog hosting, access, and updates}
The current (at time of publication) version of the catalog is released on Zenodo (DOI: 10.5281/zenodo.7894467), the RMTable package Github page, and in the {\it Vizier} catalog library, in FITS, TSV, and VOTable formats. To maximize the future value of the consolidated catalog, we intend to update it with additional catalogs on a best-effort basis as new catalogs are released. New versions of the consolidated catalog will have a unique version number and will be similarly hosted on Zenodo and Github; the full version history of the catalog will be kept within in both repositories. Zenodo will be used to produce a new citable DOI for each major version; the most recent version will also be findable through the common DOI (10.5281/zenodo.6702842). The Github repository also contains all of the individual catalogs in FITS format.

New RM catalogs (either newly published, or previously published but newly converted to RMTable2023 format) can be contributed to the consolidated catalog by contacting one of the maintainers (listed on the repository) or submitting a pull request that adds the new table to the directory of individual catalogs. The consolidated catalog will be updated with new catalogs after some quality assurance by the maintainers. When new catalogs are published without using the RMTable2023 format the maintainers may reach out to the authors about creating an RMTable2023 version of their catalogs, or may do the conversion themselves using the published tables, on a best-effort basis.

\subsection{Guidelines for catalog usage}
This consolidated catalog is a very heterogenous data set, and so some care must be exercised when using it. Below we suggest a few considerations that should be made when using values in the catalog:
\begin{itemize}
\item Sources may appear in the consolidated catalog multiple times, if they are present in multiple input catalogs, and sources reported with multiple RMs/polarized components will appear as multiple rows in the table. 
\item A small number of sources (71) have no reported errors in RM, and so may not be suitable for some types of statistical analysis. Some sources have very small errors in RM (which may or may not be justified, depending on the frequencies used in the RM determination), and one source has an RM error of exactly zero.
\item Stokes $I, Q, U$, and $V$ values cannot be directly compared between catalogs, as they will generally have different reference frequencies (which may not be known for some catalogs). For sources with reported spectral indices and reference frequencies it may be possible to estimate the Stokes $I$ values at different frequencies, but this is a small fraction of the catalog. Some sources may be variable in time, leading to different values for the same source between catalogs at different epochs.
\item Only a small number of RMs (848) have been corrected for ionospheric Faraday rotation. Most have either no correction (11 515) or are not known whether they are corrected or not (43 456), so these may have some unknown contributions from the ionosphere which may act as systematic errors within each catalog.
\item No quality assessment or vetting of values in the original catalogs has been performed, beyond requiring that the catalogs appear in published papers. Sources with unphysical parameters (e.g., negative Stokes I, fractional polarization greater than 1) have not been removed wherever they are present in the published catalogs.
\item Matching sources between catalogs is complicated by the different beam sizes of each catalog; beam size is recorded for only a small subset of catalogs. Also, some very old catalogs used few significant digits for position (as coarse as 0.1\degr), which can lead to apparent position offsets.
\item Not all RMs are statistically independent, as there have been cases of the same observations being re-processed and having new RMs determined. For example, both \citet{Brown2003} and \citet{VanEck2021} used the same observations, and \citet{Rossetti2008} included the data from \citet{Klein2003} with newer observations to re-determine the RMs. This may become more common in the future if our proposal in Section~\ref{sec:polspectra} to standardize the publishing of polarization spectra succeeds in developing a culture of re-analyzing data as additional observations or techniques become available.
\item Measurements of Faraday complexity, where present, are strongly dependent on the range of observed frequencies as to how sensitive they are to different levels of Faraday thickness/complexity \citep{Anderson2016}. Sources that appear Faraday simple/complex in one observation may not be the same in observations with different $\lambda^2$ coverage.
\end{itemize}

Users of the consolidated catalog may want to filter the catalog to remove sources not suitable for the type of analysis they are performing. Examples of several types of filters based on the considerations listed above are included in the RMTable python package.

In addition, we strongly emphasize the importance of referencing the original RM catalogs when using this consolidated catalog to find sources; citing only the consolidated catalog is not sufficient as it will reduce recognition of the value contributed by each catalog.

\section{Polarization spectra standard: PolSpectra2023}\label{sec:polspectra}
Broadband polarization spectra are critically important for Faraday rotation studies, particularly those looking for Faraday complexity. While some very early polarization studies reported the polarization properties at each of the (very few) frequencies observed \citep[e.g.,][]{Tabara1980}, most papers in recent years have focused on deriving RM values or other analysis and have not included the underlying polarization spectra. These spectra have significant scientific value beyond the individual studies they are produced for, and could be used for further studies if made available. An example of this is ultra-broadband studies that may only be possible by combining data from multiple instruments, or re-analysis of data using algorithms developed after the initial publication. With the increasing emphasis on, and usefulness of, making published data available online for later use, it is worthwhile to consider how this can be done in a way that is efficient and easy to use. Here we propose a format for storing and manipulating polarized spectra of compact sources.

A difficulty with storing or archiving polarization data sets is that such data are generally four dimensional, position-position-frequency-Stokes, which results in high data volumes. Since the polarized sky is more sparsely populated than the total intensity sky, most of the pixels are not needed. For sources that are unresolved or without interesting resolved structure (i.e., those within the scope of the catalog standard of Sec.~\ref{sec:standard}), the position axes can be dropped, reducing the data to a set of one dimensional frequency spectra with a corresponding significantly reduced volume. However, since the pixels adjacent to the source are often used to estimate the uncertainties in the source intensity or flux density, this information would be lost in this process, so it would be useful to also create and keep `uncertainty spectra' for each Stokes parameter alongside the sources' spectra.

We considered several alternatives for structuring and storing the spectrum data, judging them on criteria including data volume efficiency, ease of including and accessing source-based metadata, ability to combine different spectra together, and ease of reading, writing, and searching the spectra files.

The first option we considered was single-pixel FITS images: maintaining the usual structure and axes of a FITS frequency cube, but clipping the position axes down to a single pixel containing the source. Each Stokes parameter would be a separate file (along with the uncertainty spectra), leading to 8 files per source.\footnote{These could be combined as separate extensions in a single FITS file, but this does not significantly mitigate the disadvantages.} The advantages to this scheme are that it would preserve all the FITS header information from the original data, as well as be easy to read and manipulate with existing FITS readers/viewers. The disadvantages are that it leads to many small files, which may cause problems for file systems if there are many sources; the inability to combine data (other than grouping files together); the inability to handle irregular sampling in frequency; and a high degree of redundancy as each file gets a complete FITS header.

Another option would be the format used by \citet{Farnes2014}, who split the data into two tables: the first containing the source meta-data (identifier, coordinates, etc; one row per source), and the second containing the channel wavelengths and polarization properties (one row per source-channel). This format had the advantages that the tables can be stored in any conventional format, and that it is easy to add new data. The disadvantages are that accessing the spectra becomes more complex (cross-referencing through two different tables) and there is some storage overhead to set up the cross-referencing.

The third option, which we have selected as the best choice, is to use tables with array columns containing the spectra. The FITS standard supports columns that contain arrays, and also supports those arrays having different lengths in different rows \citep{Cotton1995}. This format has the advantages of being easy to read, write, and manipulate both spectra and metadata, relatively efficient for storage volume, and allows different observations to be combined into the same file. The disadvantage is that some amount of metadata may be repeated, reducing storage efficiency, but this can be countered by using standard file compression algorithms. We also note that all of these advantages apply to using the same structure with VOTables.

Below we propose a standard for a tabular format for storing polarization spectra, which we call `PolSpectra2023', and describe a Python module we have designed to read and write these tables as FITS or VOTable files. It is not our intention to maintain a consolidated database of polarization spectra, in contrast with the RM catalog of Section~\ref{sec:catalog}; instead this standard is intended to serve as a suggestion for those interested in publishing polarization data in a way that maximizes ease of use and compatibility between datasets. 

\subsection{Proposed standard}
The PolSpectra2023 standard is intended to have the same scope as the RMTable2023 standard: for radio sources or source components that can be reasonably characterized using only a single spectrum. Well-resolved sources with spatially varying structure are beyond the intended use of this standard, although there is nothing preventing it from being used to store integrated flux density spectra of such sources. The standard retains no information on spatial structure of a source,\footnote{The standard does not prevent users from attaching spatial information, such as deconvolved sizes, as additional columns, but it does nothing to explicitly encourage this.} as it is intended to target cases where such information is not present (unresolved sources) or unneeded for some kinds of analysis; cases where the full spatial data is needed are already well handled by existing data-cube products (e.g., FITS image cubes).

Each row of a PolSpectra2023 table is a single source (or source component, for sources resolved into multiple individual spatial components) observation; separate observations (in different frequency bands) of the same source appear as separate rows in the table. This allows for observation-dependent metadata associated with a source to be kept when multiple observations are combined. A source number column is included in the standard to make it easier to associate data belonging to same source, allowing users to select all observations of the same source using the source number rather than relying on more complicated coordinate cross-matching.

Similarly to the RMTable2023 standard, PolSpectra2023 has a defined set of required columns that must be present, and a set of suggested optional columns; these are summarized in Table~\ref{tab:PolSpectra}. Users creating PolSpectra2023 tables are not limited to only these columns, but columns that are explicitly part of the standard should use the standard names and units. This is required to ensure the ability to straightforward combine and interact with tables without issues from incompatibilities between tables. Short descriptions of all of the defined columns are given below. Similarly to the RMTable2023 standard, it is not specified whether floating point numbers should be 32- or 64-bit (except for coordinates, which should be 64-bit), and any of the required columns without entries should have values of NaN. All columns specified as arrays should have the same number of entries in each row as the corresponding frequency array for that row.

\startlongtable
\begin{deluxetable*}{lllccll}
\tablecaption{Column definitions for the PolSpectra2023 standard.\label{tab:PolSpectra}}
\tablehead{\colhead{Column Name} & \colhead{Short description} &  \colhead{Data format} & \colhead{Unit} & Limits & Default/Missing}
\startdata
\multicolumn{6}{c}{\bf Required columns}\\
source\_number & ID number for the source & integer & -- & -- & Essential \\
ra & Right ascension (ICRS) & double & deg & [0,360) & Essential \\
dec & Declination (ICRS) & double & deg & [-90,90] & Essential  \\
l & Galactic Longitude & double & deg & [0,360) & Essential \\
b & Galactic Latitude & double & deg & [-90,90] & Essential \\
freq & Channel frequencies & float array & Hz & (0,$\infty$) & Essential \\
stokesI & Stokes $I$ intensities or flux densities & float array & Jy or Jy/beam & (-$\infty$,$\infty$) & NaN \\
stokesI\_error & uncertainties in Stokes $I$ per channel & float array & Jy or Jy/beam & (0,$\infty$) & NaN \\
stokesQ & Stokes $Q$ intensities or flux densities & float array & Jy or Jy/beam & (-$\infty$,$\infty$) & NaN \\
stokesQ\_error & uncertainties in Stokes $Q$ per channel & float array & Jy or Jy/beam & (0,$\infty$) & NaN\\
stokesU & Stokes $U$ intensities or flux densities & float array & Jy or Jy/beam & (-$\infty$,$\infty$) & NaN \\
stokesU\_error & uncertainties in Stokes $U$ per channel & float array & Jy or Jy/beam & (0,$\infty$) & NaN \\
beam\_maj & Major axis of the synthesized beam & float or float array & deg& (0,$\infty$) & NaN \\
beam\_min & Minor axis of the synthesized beam & float or float array & deg & (0,$\infty$) & NaN\\
beam\_pa & Position angle of the synthesized beam & float or float array & deg & (0,$\infty$) & NaN \\
Nchan & Number of channels & int & -- & [1,$\infty$) & Essential \\
\multicolumn{6}{c}{\bf Suggested/optional columns:}\\
stokesV & Stokes $V$ intensities or flux densities & float array & Jy or Jy/beam & (-$\infty$,$\infty$) & NaN \\
stokesV\_error & uncertainties in Stokes $V$ per channel & float array & Jy or Jy/beam & (0,$\infty$) & NaN  \\
quality & Data quality flags per channel & int array & -- & -- & -- \\
quality\_meanings & Short description of quality flags & str & -- &-- & -- \\
ionosphere & Ionospheric correction method & str & -- & -- & -- \\
cat\_id & Source name/ID in data paper & str & -- & -- & -- \\
dataref & Reference to data paper & str & -- & -- & -- \\
telescope & Name of Telescope(s) & str & -- & -- & -- \\
epoch & Observation epoch (midpoint, MJD) & float & days & (-$\infty$,$\infty$) & NaN \\
integration\_time & Integration time of observations & float & s &  (0,$\infty$) & NaN \\
interval & Interval of observation & float & days & [0,$\infty$) & NaN \\
leakage & Instrumental leakage estimate & float or float array & -- & [0,$\infty$) & NaN \\
channel\_width & Bandwidth of each channel & float or float array & Hz & (0, $\infty$) & NaN \\
flux\_type & Stokes extraction method & str & -- & -- & -- \\
aperture & Integration aperture for spectra& float & deg & [0, $\infty$) & NaN \\
\enddata
\end{deluxetable*}

\subsubsection{Source ID number}
An integer that uniquely identifies the source in the table, in order to group together different observations of the same source. These numbers are unique to each table and in general should not be preserved under table merging operations; care must be taken when combining different PolSpectra2023 tables together to ensure different sources are not accidentally grouped together. In cases where it may be ambiguous how to assign such groupings (e.g., a source is resolved/divided into multiple components in one observation but not another), it is left to the discretion of the table author how to address the problem.

\subsubsection{Right ascension and declination, Galactic longitude and latitude}
The source coordinates, in the equatorial (ICRS) and Galactic frames, in decimal degrees, are required columns. Double precision (64-bit) is required if the source positions are more accurate than a 32-bit float can accommodate. Having both coordinate systems is preferred for ease of selecting populations in either coordinate system as desired.

\subsubsection{Channel frequencies}
An array containing the center frequencies of all channels in the spectra, in Hz. All other channel-based columns must have the same number of entries. Most polarimetric data sets report the frequency in the topocentric (observatory) reference frame, so exceptions to this should be explicitly documented.

\subsubsection{Stokes parameters and uncertainties}
For each Stokes parameter, an array containing the values for all channels (as floats). Each Stokes parameter also has a corresponding column containing an array of uncertainties. Stokes $I$, $Q$, and $U$ are required (both values and uncertainties), Stokes $V$ is optional, but strongly encouraged. All values must be either intensities, in Jy/beam, or flux densities, in Jy. Stokes $I$ is not required to be greater than zero, since negative values can occur due to noise in individual channels.

\subsubsection{Beam size and orientation}
Three required columns that define the major axis FWHM, minor axis FWHM, and position angle of the synthesized beam (assuming a Gaussian beam model), all in degrees. Position angle is measured east from north. Each of these columns may be either scalar floats (implying the same beam shape for all frequencies) or float arrays if the beam is frequency-dependent. These are required to allow users to understand the effects of the beam on the extracted spectra.

\subsubsection{Number of channels}
The number of channels in the observation, as an integer. Must be equal to the size of the arrays used in the channel-dependent columns; this column is used for validating the spectra by checking that all channel columns have this number of values.

\subsubsection{Channel quality flags and flag meanings}
Two optional columns that can contain information about the quality of the Stokes spectra per-channel. The quality flag column contains an array of integers that encode information about the quality of the channels. The quality meanings column contains a string that provides a short explanation for each flag value. For example, this string could be ``0=good channel; 1=high-noise; 2=flagged''.

\subsubsection{Ionospheric correction method.}
A string describing the method used to correct the Stokes $Q$ and $U$ spectra for ionospheric Faraday rotation, if any. If no correction has been applied, this should contain `None', otherwise this should take the form of the software name, a paper reference, or short algorithm name. Suggested values for common/previously used methods are given in Appendix~\ref{app:standards} and will be updated in the online documentation as needed. This column is useful for informing the user about the possible presence (or removal) of ionospheric Faraday rotation which may act as a systematic bias or error in analyses using the spectra.

\subsubsection{Source name/ID}
A short string containing the name or catalog ID assigned to the source in the paper where the spectra were published.

\subsubsection{Data reference}
A string containing reference(s) to the paper(s) in which the spectra were published or described. The preferred format for this is an ADS bibcode, or the DOI. If the data were published separately from a publication with a specific DOI, both the data DOI and a reference to the paper in which the data are described should be included. If multiple references are included, a comma-separated list should be used.

\subsubsection{Name of telescope(s)}
A string containing the names or acronyms of all telescopes that contributed to the spectra, as a comma separated list when more than one is present. Note that it is intended that where different parts of the spectra are observed by different telescopes, they should be reported as different rows (in order to properly propagate/preserve observation data such as epoch or beam shape). Only in cases where multiple telescopes were used to produce a measurement (e.g., combining single dish and interferometer data) should all those telescopes be included in the same row. A list of currently used or suggested names appears in Appendix~\ref{app:standards} and will continue to be updated in the online documentation as new values come into use, in order to encourage authors to use the same values when using the same methods.

\subsubsection{Epoch, integration time, and interval of observation}
Three optional columns that contain the median epoch of observation (in MJD), the integration time of the observation (in seconds), and the interval of observation (time between first observation and last observation, in days). This information can be very useful for studies looking at time-variability or evolution of sources.

\subsubsection{Instrumental leakage estimate}
An estimate of the degree of instrumental leakage in Stokes $Q$ and $U$, as a fraction of Stokes $I$. This can either be a single floating point value (assumed to apply to all channels), or an array of floats giving the estimate for each channel. If a leakage correction has been applied, this should be an estimate of the residual leakage. This estimate can be helpful for users who wish to understand the presence of systematic errors in the data.

\subsubsection{Channel bandwidth}
An optional column that gives the bandwidth of the channels. This can either be a single (float) value, if all the channels have the same bandwidth, or it can be an array of floats giving the bandwidth of each channel. While channel bandwidth can often be inferred from the separation of channels, it can be beneficial to have it explicitly given for users concerned about the effects of bandwidth depolarization.

\subsubsection{Stokes extraction method}
A string describing the method used to extract the source spectra, for example whether they were intensities derived from peak-pixel values, aperture-integrated flux densities, or intensities or flux densities derived from point-source or Gaussian fitting. If the method is not known, the default value is `Unknown'. A list of currently used or suggested values appears in Appendix~\ref{app:standards} and will continue to be updated in the online documentation as new values come into use, in order to encourage authors to use the same values when using the same methods. This column can be useful for users to understand the process by which spectra values were derived, which can affect comparisons with other data as well as understanding results from (partially) resolved sources or sources with multiple components.

\subsubsection{Integration aperture}
An optional column that gives the linear size of the integration aperture (as a float) over which the spectra have been integrated or averaged, in degrees. If only peak/single pixel values are extracted from the images, this should be zero. If a Gaussian-fit or similar process was used, then the FWHM of the fitted area would be appropriate. This information, in combination with the Stokes extraction method, should provide enough information to reproduce a spectrum extraction from the original data.

\subsection{PolSpectra module}
We have developed a package in Python 3, called PolSpectra, for the creation and manipulation of tables following the above standard. This package supplies functions that allow a user to build PolSpectra2023 tables and manipulate them using standard numpy and Astropy Tables methods, write them to FITS binary table and VOTable formats, perform simple merging of tables, and read PolSpectra2023 tables into Python. Appendix~\ref{app:Spectra_example} shows examples of how this package can be used. The code for this package is available through Github\footnote{https://github.com/CIRADA-Tools/PolSpectra} and PyPI.

\section{Summary and discussion}\label{sec:summary}
Rotation measure catalogs of polarized radio sources, and the polarized spectra from which they are derived, are useful for a broad range of astronomical studies, so it is valuable to ensure that these data are as easy as possible to access and work with. With this goal, we have proposed two data standards that can be used when publishing radio polarization data. The first, `RMTable2023', defines a set of parameters that are beneficial to include in RM catalogs in order to maximize potential value for future studies. The second, `PolSpectra2023', defines a set of parameters and a format for storing full Stokes spectra of radio sources, in order to make those spectra (more easily) accessible to other researchers.

To enhance and demonstrate the value of the RMTable2023 standard, we have assembled a consolidated catalog of previously published RMs, converted into an RMTable, containing 55 819 RMs. This catalog is the largest assembled to date, and should significantly simplify the task of finding sources with measured RMs in any particular region of sky or fitting some specific parameters. However, this comes with the caveat that the individual catalogs from which these RMs are drawn are very heterogenous, and in many cases some source properties or observation parameters are not easily available. We intend to extend and update this catalog as resources permit, so authors of new RM catalogs are encouraged to submit new catalogs through email or Github pull requests; we also invite the community to contribute to converting older catalogs into RMTable2023 format for inclusion in the consolidated catalog.

We propose that researchers who publish results derived from polarized spectra of compact radio sources, and/or RMs determined from such data, should release their data using these two data standards in order to maximize the value and ease-of-use of these data for future research. This will vastly simplify scientific projects such as re-analyzing these data with new or different algorithms, combining multiple datasets to enable new analyses, and will increase the power of statistical studies requiring large samples of sources. These PolSpectra and RMTable data can be archived online using a variety of services, including the publishing journal, the arXiv, Vizier, and Zenodo. A few RM catalogs have already published in the RMTable2023 format \citep{Riseley2020,Ma2020,VanEck2021,Livingston2021}, and the POSSUM, VLASS, SPICE-RACS, and Apertif surveys will make their polarization data available in PolSpectra2023 and RMTable2023 format, so we expect there to soon be a large quantity of data available in these formats. Publishing data using these standards should not cause a significant burden on the authors of new polarization data; the time investment is expected to be at the level of only a few hours for most catalogs, especially if planned for early in the data reduction process.

We intend to maintain and update the consolidated RM catalog for as long as feasible, adding newly published RMs as they are released; this effort will also be greatly eased by having these RMs published following the RMTable2023 standard. Updated versions of the catalog will be published periodically, and can be found through the GitHub repository for the RMTable package. We emphasize again that this catalog is intended as a resource for users to easily find RMs relevant to their research, it is important for catalog users to cite the publications from which these RMs were found. We do not plan to maintain a single central collection for PolSpectra2023 data, but a list of published data with links to their archival locations will be kept with the documentation for the PolSpectra package. The development of these standards and the consolidated catalog may be an early step towards the development of a radio polarization-oriented Virtual Observatory that could serve as a clearing house for such data in the future.

\acknowledgments
We would like to thank the radio polarimetry community for their feedback in the early stages of organizing this work, and both referees for their thoughtful and thorough comments.

This work has made extensive use of the following software packages: Astropy, a community-developed core Python package for Astronomy \citep{Astropy,Astropy2}; SciPy \citep{Scipy}; NumPy \citep{Numpy}; and Matplotlib \citep{Matplotlib}. This research has made use of NASA’s Astrophysics Data System, and the Vizier catalog service operated by the Centre de Données astronomiques de Strasbourg.

The Canadian Initiative for Radio Astronomy Data Analysis (CIRADA) is funded by a grant from the Canada Foundation for Innovation 2017 Innovation Fund (Project 35999) and by the Provinces of Ontario, British Columbia, Alberta, Manitoba and Quebec, in collaboration with the National Research Council of Canada, the US National Radio Astronomy Observatory and Australia’s Commonwealth Scientific and Industrial Research Organisation. The Dunlap Institute is funded through an endowment established by the David Dunlap family and the University of Toronto.

CJR acknowledges financial support from the ERC Starting Grant `DRANOEL', number 714245. SH acknowledges funding from the European Research Council (ERC) under the European Union’s Horizon 2020 research and innovation programme (grant agreement No. 772663). BA acknowledges funding from the German Science Foundation DFG, within the Collaborative Research Center SFB1491 ''Cosmic Interacting Matters - From Source to Signal''.

\bibliography{References}{} 
\bibliographystyle{aasjournal} 

\appendix
\section{Lists of suggested standard values}\label{app:standards}
Several of the columns in the RMTable2023 and PolSpectra2023 standards are strings that are intended to contain information about that data, such as the telescope(s) used or methods used in processing. To make it easier for these columns to be understood, searched, or used in row-selection operations, we propose standard values for these quantities, which are given in tables below; these tables contain values that appear in the consolidated catalog or are expected to appear in upcoming catalogs. These are not intended to be prescriptive, but to encourage consistency across as many catalogs as possible. We encourage authors creating new catalogs to use these values when applicable, while recognizing that these are not comprehensive. Updated versions of these tables will be maintained online with the RMTable and PolSpectra code packages, with new values added as they come into use in new catalogs.

\subsection{RM determination method}
The values in Table~\ref{tab:rm_method} are suggested for the `rm\_method' column of RMTable2023.

\begin{table}[h]
\caption{Standard values for RM determination method}\label{tab:rm_method}
\begin{center}
\begin{tabular}{|l|l|} \hline
{\bf Standard value:} & {\bf Notes:} \\ \hline
Unknown & Default value if not specified.\\
EVPA-linear fit & Linear regression of polarization angle as function of wavelength-squared.\\
RM Synthesis & Any variation of the RM-synthesis algorithm\\
RM Synthesis - Pol. Int & RM synthesis performed using measured Stokes $Q$ and $U$ values.\\
RM Synthesis - Fractional polarization & RM synthesis performed using $Q$ and $U$ values normalized by Stokes $I$.\\
Faraday Synthesis & Joint aperture and RM synthesis, as described in \citet{Bell2012}.\\
QUfit & Any variation of QU-fitting algorithm.\\
QUfit - Delta function & QU-fitting of a Faraday-simple model.\\
QUfit - Burn slab & QU-fitting of a slab model from \citet{Burn66}.\\
QUfit - Gaussian & QU-fitting of a Gaussian model.\\
QUfit - Gaussian x Burn Slab & QU-fitting of a model which is the product of a Gaussian and Burn slab.\\
QUfit - Multiple & QU-fitting to a model that is a combination of different models.\\
\hline
\end{tabular}
\end{center}
\label{default}
\end{table}%

\subsection{Faraday complexity metric}
The values in Table~\ref{tab:complex_test} are suggested for the `complex\_test' column of RMTable2023.

\begin{table}[h]
\caption{Standard values for Faraday complexity metric}\label{tab:complex_test}
\begin{center}
\begin{tabular}{|l|p{0.6\linewidth}|} \hline
{\bf Standard value:} & {\bf Notes:} \\ \hline
`'  (empty string) & Default value if not specified.\\
None & No complexity test performed.\\
Sigma\_add & The $\sigma_\mathrm{add}$ method implemented in RM-Tools \citep{RM-Tools}.\\
Second\_moment & Analysis of the second moment of RM-clean components.\\
QU-fitting & QU-fitting of Faraday complex model.\\
Inspection & Visual inspection of FDF for deviations from Faraday-simple response.\\
Machine learning - Alger 2021 & Machine learning algorithm developed by \citet{Alger2021}.\\
Convolutional neural networks - Brown 2019 & Convolutional neural network algorithm developed by \citet{Brown2019}.\\
QU-fit \& BIC & Bayesian Information Criterion applied to QU-fitting model (comparing simple vs complex models). \\
 \hline
\end{tabular}
\end{center}
\label{default}
\end{table}%

\subsection{Ionospheric correction method}
The values in Table~\ref{tab:ionosphere} are suggested for the `ionosphere' column of RMTable2023 and PolSpectra2023.

\begin{table}[h]
\caption{Standard values for ionospheric Faraday rotation  correction method}\label{tab:ionosphere}
\begin{center}
\begin{tabular}{|l|l|} \hline
{\bf Standard value:} & {\bf Notes:} \\ \hline
Unknown & Default value if not specified.\\
None & No ionospheric correction performed.\\
RMextract & The RMextract package \citep{RMextract}.\\
ionFR & The ionospheric Faraday rotation package \citep{Sotomayor13}. \\
FARAD & The AIPS task FARAD.$^a$\\
ALBUS & The ionosphere correction tool in the Advanced Long Baseline User Software (ALBUS).$^b$\\
FRion & The FRion Python package for time-averaged image-domain ionospheric correction.$^c$\\
\hline
\end{tabular}
\tablecomments{
$a$. www.aips.nrao.edu/cgi-bin/ZXHLP2.PL?FARAD\\
$b$. github.com/twillis449/ALBUS\_ionosphere\\
$c$. frion.readthedocs.io
}
\end{center}
\label{default}
\end{table}%

\subsection{Polarization bias correction}
The values in Table~\ref{tab:pol_bias} are suggested for the `pol\_bias' column of RMTable2023. These values are generally the bibcode of the paper where the correction equation was defined.

\begin{table}[h]
\caption{Standard values for polarization bias correction method}\label{tab:pol_bias}
\begin{center}
\begin{tabular}{|l|l|} \hline
{\bf Standard value:} & {\bf Notes:} \\ \hline
Unknown & Default value if not specified.\\
None & No bias correction performed.\\
Not described & The paper reports that a correction was performed, but does not specify which method or equation.\\
1974ApJ...194..249W & \citet{Wardle1974}\\
1985A\&A...142..100S & \citet{Simmons1985}\\
1986ApJ...302..306K & \citet{Killeen1986}\\
2012PASA...29..214G & \citet{George2012}\\
\hline
\end{tabular}
\end{center}
\label{default}
\end{table}%

\subsection{Telescope}
The values in Table~\ref{tab:telescope} are suggested for the `telescope' column of RMTable2023 and PolSpectra2023. These values can be combined using semicolons if multiple telescopes have contributed to the data.

\begin{table}[h]
\caption{Standard values for telescope used}\label{tab:telescope}
\begin{center}
\begin{tabular}{|l|l|} \hline
{\bf Standard value:} & {\bf Notes:} \\ \hline
Unknown & Default value if not specified.\\
VLA & Karl G. Jansky Very Large Array\footnote{To avoid splitting entries between VLA and JVLA, we suggest only using VLA.}\\
LOFAR & Low Frequency Array\\
ATCA & Australia Telescope Compact Array\\
DRAO-ST & Dominion Radio Astrophysical Observatory Synthesis Telescope\\
MWA & Murchison Widefield Array\\
ATA & Allen Telescope Array\\
WSRT & Westerbork Synthesis Radio Telescope\\
ASKAP & Australian Square Kilometre Array Pathfinder\\
Effelsberg & Effelsberg 100-m Radio Telescope\\
ARO & Algonquin Radio Observatory 46-m telescope\\
MeerKAT & \\
Arecibo & Arecibo Telescope\\
Parkes & Parkes Murriyang (64-m) Radio Telescope\\
CHIME & Canadian H Intensity Mapping Experiment\\
FAST & Five-hundred-meter Aperture Spherical Telescope\\
\hline
\end{tabular}
\end{center}
\label{default}
\end{table}%

\subsection{Stokes extraction method}
The values in Table~\ref{tab:flux_type} are suggested for the `flux\_type' column of RMTable2023 and PolSpectra2023. 

\begin{table}[h]
\caption{Standard values for Stokes extraction method}\label{tab:flux_type}
\begin{center}
\begin{tabular}{|l|l|} \hline
{\bf Standard value:} & {\bf Notes:} \\ \hline
Unknown & Default value if not specified.\\
Peak & Stokes values extracted from single pixel (peak Stokes I or polarized intensity).\\
Integrated & Stokes values integrated over some aperture.\\
Gaussian fit - Peak & Stokes values from peak of fitted Gaussian.\\
Gaussian fit - Integrated & Stokes values from integrated brightness of fitted Gaussian.\\
Box & Stokes values determined from integration over fixed box size.\\
Visibilities & Stokes values extract from modelling or fitting of interferometric visibilities.\\
\hline
\end{tabular}
\end{center}
\label{default}
\end{table}%

\subsection{Source classification}
The values in Table~\ref{tab:source_type} are suggested for the `type' column of RMTable2023.

\begin{table}[h]
\caption{Standard values for source classification}\label{tab:source_type}
\begin{center}
\begin{tabular}{|l|l|} \hline
{\bf Standard value:} & {\bf Notes:} \\ \hline
`' (empty string) & Default value if not specified.\\
Unknown & \\
Pulsar & \\
Galaxy & Radio Galaxy \\
AGN & Active Galactic Nucleus \\
SNR & Supernova Remnant\\
FRB & Fast radio burst\\
\hline
\end{tabular}
\end{center}
\label{default}
\end{table}%

\section{Notes on specific catalogs within the consolidated RM catalog}\label{app:papers}
Some of the individual catalogs within the consolidated RM catalog required some changes in order to be incorporated, or have some specific details that may influence how users may want to use those RMs. We list those details here, ordered from largest catalog to smallest. Trivial changes, such as converting coordinates from sexagesimal to decimal degrees, unit changes, or including information explicitly given in the paper accompanying each catalog, are not described here.

{\bf \citet{Taylor09}:} The 1D position error was calculated for each source as the larger of the (de-projected) RA error or the declination error.

{\bf \citet{Schnitzeler2019}:} While the authors report fitting up to 5 polarized components per source, the available data table only reports the brightest 2 components per source. These two components have been incorporated into the consolidated catalog (as separate rows), and the number of components column has been capped at two for these sources. The 1D position error was calculated for each source as the larger of the (de-projected) RA error or the declination error. We also note that some sources have questionable values for some columns such as negative Stokes I values or unphysical spectral indices or fractional polarizations; these have been left unchanged.

{\bf \citet{Brown2003, VanEck2021}:} These catalogs used the same data, so RMs present in both catalogs are not independent measurements.

{\bf \citet{Farnes2014}:} The RMs in this catalog were determined using data from multiple sources, some of which were also used to determine RMs. As a result, these RMs are not fully independent measurements of the same sources that are also present in the \citet{Klein2003}, \citet{Rossetti2008}, and \citet{Taylor09} catalogs. To incorporate this catalog we constructed a look-up table to work out which data references, telescopes, and frequencies were used for each source.

{\bf \citet{Tabara1980}:} This catalog is a collection of published RMs up to December 1978. The process of determining the original publication on a per-RM basis was deemed too difficult to be done when incorporating this catalog into the consolidated catalog.

{\bf \citet{Broten1988}:} This catalog is a collection of published RMs from 10 separate papers published from 1975 to 1988. As best as we can determine, there is no duplication of entries with \citet{Tabara1980}. As above, finding the original publication for each source was deemed too difficult. No direct machine-readable version of this table was found online, so the RMs were incorporated from a table compiled by Jo-Anne Brown. The position of each RM is precise to only 0.1\degr\ (in $l$ and $b$) due to the limited significant figures used in this table.

{\bf \citet{Simard-Normandin1981}:} The RMs in this catalog were determined in part using previously published measurements; it is not clear if there is any overlap with the data used for the RMs in the \citet{Tabara1980} and \citet{Broten1988} catalogs. The position of each RM is also limited in precision to 0.1\degr\ due to significant figures.

{\bf \citet{Riseley2018, Riseley2020}:} These catalogs use some of the same data as each other, so sources present in both catalogs may not be fully independent measurements.

{\bf \citet{OSullivan2017}:} The {\it Vizier} version of this catalog does not include the measured RMs; the RM values were extracted from the paper's LaTeX source available on the arXiv. This catalog reports multiple components for some sources; each component was converted to a separate row in the consolidated catalog. The RM width column was determined by combining the $\sigma_\mathrm{RM}$ and $\Delta_\mathrm{RM}$ columns of the original tables (each component had a value for only one or the other). Sources fit with a Faraday thin model have NaN values for RM width.

{\bf \citet{Klein2003}:} The data these authors used to determine RMs have also been used in other projects \citep{Rossetti2008, Farnes2014}, so these RMs may not be statistically independent.

{\bf \citet{Heald09}:} Some of the sources in this catalog are resolved and have RMs determined for different (spatially-separated) components, but the catalog does not provide coordinates for each component. In these cases, all the components have been given the same coordinate.

{\bf \citet{Clarke2001}:} The RM values are not available online, but were supplied by the author.

{\bf \citet{Clegg1992, Oren1995, Minter1996}:} No online machine-readable version of these catalogs were found; these RMs were taken from a table assembled by Jo-Anne Brown. The position of each RM is limited in precision to 0.1\degr\ due to significant figures.

{\bf \citet{Battye2011}:} The source positions are based on short names, resulting in a position accuracy limited to approximately 0.1\degr.

{\bf \citet{Costa2018}:} The authors supplied RMs calculated from two different methods; we have used the EVPA linear fitting method values for the RM column in the consolidated catalog.

\section{Example usage of RMTable package}\label{app:RMTexample}
Below we give a few examples of how to interact with an RMTable object in Python 3, using the RMTable package described in Section~\ref{sec:RMTable}. The first example demonstrates read/write operations, and simple catalog exploration/manipulation. These examples are also present on the package Github repository.

\begin{lstlisting}[language=Python]
import numpy as np
from rmtable import RMTable
from astropy.coordinates import SkyCoord


#Reading in an RMTable from FITS:
catalog=RMTable.read('VanEck2011_table.fits')
print(catalog)

#Get the list of columns present in table:
print(catalog.columns)

#Get number of RMs (2 methods):
print(catalog.size)
print(len(catalog))

#Access column(s):
print(catalog['rm'])
print(catalog['l','b','rm','rm_err'])

#Access row(s):
print(catalog[0:10])

#Extracting a subset of the catalog:
selection=np.logical_and(catalog['l'] > 90,catalog['l'] < 270)
#This is an array of booleans, that can be used to 
#extract a portion of the catalog.
print(catalog[selection])
#Multiple selections are combined using numpy's logical_and function.


#If you prefer numpy arrays or pandas dataframes, convert to those
print(catalog.as_array())
print(catalog.to_pandas())


#Saving a sub-catalog:
subcatalog=catalog[selection]
subcatalog.write('subcatalog.fits',overwrite=True)
#---------------------------------------------------------#
\end{lstlisting}

The second example demonstrates how a machine-readable table of an existing RM catalog can be converted to an RMTable.

\begin{lstlisting}[language=Python]
import numpy as np
from rmtable import RMTable
import astropy.coordinates as ac

#How to convert other tables containing RMs into RMTables.
#Read in machine-readable table (fixed-width ASCII file) of catalog,
#into numpy ndarray. Note the fixed width columns are set with the delimiter 
#keyword. Columns that match the standard are given names in the standard, 
#to allow direct conversion; other columns must avoid name conflicts with
#standard columns.
cat=np.genfromtxt('VanEck2011.dat',encoding=None,dtype=None,
                  delimiter=[6,6,3,3,5,2,2,3,3,5,4,5,3,5,3,7,5],
                    names=['l','b','rah','ram','ras','dec_sign','decd',
                           'decm','decs','stokesI','polint','rm','rm_err',
                           'RMSynth','dRMsynth','NVSSRM','dNVSSRM'])
#Setting the column names correctly is important to get the data into the RMTable.
# Columns with incorrect names are ignored in the conversion process.



#The RA and Dec columns must be converted from sexigessimal to decimal.
#The easiest way is to use Astropy's capability to read 'hms dms' strings:
ra_strings=np.char.add(np.char.add(np.char.add(cat['rah'].astype(str),'h'),
                                   np.char.add(cat['ram'].astype(str),'m')),
                                   np.char.add(cat['ras'].astype(str),'s'))
dec_strings=np.char.add(cat['dec_sign'],
                np.char.add(np.char.add(np.char.add(cat['decd'].astype(str),'d'),
                                        np.char.add(cat['decm'].astype(str),'m')),
                                        np.char.add(cat['decs'].astype(str),'s')))
coords=SkyCoord(ra_strings,dec_strings,frame='fk5')
#Adding final decimal coordinate columns in to the numpy table:
cat=np.lib.recfunctions.append_fields(cat,['ra','dec'],
        [coords.ra.deg,coords.dec.deg])


#Step 2: do necessary unit conversions to match RMTable convention.
#In this example, converting fluxes from mJy in the input table to 
#Jy in the RMTable.
cat['polint']=cat['polint']/1e3
cat['stokesI']=cat['stokesI']/1e3

#Step 3: convert to RMTable. It will automatically identify which columns are
#        part of the standard and which are not, based on the column names.
##table = RMTable(cat)
table=RMTable.input_numpy(cat,verbose=True,verify=True,coordinate_system='fk5')
#Tt will report which columns were used or ignored, and which
#are missing and filled with blanks.
#If verify=True, it will check that the numerical values are as expected.
#This typically means things like angle conventions (i.e. polarization angles 
#from [0,180)).
#The coordinate system must be specified to ensure that coordinates are 
#successfully converted to ICRS.


#Step 4: add any information that wasn't in the input table (but is in the text
#        of the paper. Most important is the catalog bibcode.
table['catalog_name']='2011ApJ...728...97V'
table['rm_method']='EVPA-linear fit'
table['ionosphere']='None'
table['flux_type']='Peak'
table['beam_maj']=0.01388888889
table['minfreq']=1365e6
table['maxfreq']=1515e6
table['channelwidth']=3.57e6
table['Nchan']=14
table['noise_chan']=2e-3
table['int_time']=120

#The following lines check the table values for conformance with the standard
# (within limits, and using standard string values where applicable).
table.verify_columns()
table.verify_limits()
table.verify_standard_strings()

#The package automatically adds all of the standard columns, filling any
#columns not supplied with the default values:
print(table['rmsf_fwhm'])
print(table['dataref'])


#Step 5: Save the RMTable. All formats supported by astropy are supported.
table.write('VanEck2011_table.fits',overwrite=True)


#---------------------------------------------------------#
\end{lstlisting}

The above code can be adapted to convert other catalogs into RMTable2023 format. The key steps, and most common potential pitfalls, are:
\begin{itemize}
\item Read the catalog table into a numpy array with named columns. The column names must match (or be changed to) the standard column names in order to be automatically converted. Care should be taken to ensure that all columns have been read in properly.
\item If the coordinates are in sexagesimal, they must be converted to decimal degrees; particular attention should be paid to ensure that sources with negative declination are processed correctly (particularly those with -1\degr\ $< \delta <$ 0\degr).
\item The table columns should be converted to the correct units and angle conventions, as necessary.
\item Convert the numpy array into an RMTable object.
\item Add information not present in the tables into the RMTable object.
\item Verify that the table values conform to the standard.
\item Save the RMtable to a file.
\end{itemize}

\section{Example usage of PolSpectra package}\label{app:Spectra_example}
The below code demonstrates how a PolSpectra2023 table can be assembled the Python 3 package described in Section~\ref{sec:polspectra}, how such a table can be manipulated in Python, and how it can be written to/read from a FITS file. These examples can also be found on the package Github repository.

\begin{lstlisting}[language=Python]
import numpy as np
import polspectra

#The following function can be used to generate fake example data for
# Stokes Q and U.
def Faraday_thin_complex_polarization(freq_array,RM,Polint,initial_angle):
    """Function to produce Stokes Q/U spectra for Faraday simple source.
       freq_array = channel frequencies in Hz
       RM = source RM in rad m^-2
       Polint = polarized intensity in whatever units
       initial angle = pre-rotation polarization angle (in degrees)"""
    l2_array=(299792458./freq_array)**2
    Q=Polint*np.cos(2*(np.outer(l2_array,RM)+np.deg2rad(initial_angle)))
    U=Polint*np.sin(2*(np.outer(l2_array,RM)+np.deg2rad(initial_angle)))
    return np.squeeze(np.transpose(Q+1j*U))

#Set up channel frequencies for all sources.
N_chan=288
freq_arr=np.linspace(800e6,1088e6,num=N_chan)

N_spectra=100
#Creating random spectra. This produces lists of arrays for each Stokes parameter.
I_spectra=[]
Q_spectra=[]
U_spectra=[]
noise_amplitude=0.001
for i in range(N_spectra):
    #Stokes I spectra are power laws with random brightnesses.
    Ivalues=np.random.lognormal(mean=-2,
    				sigma=1,size=N_chan)*
				(freq_arr/np.mean(freq_arr))**-0.7
    polarization=Faraday_thin_complex_polarization(freq_arr,
    				RM=np.random.uniform(-1000,1000), #Random RM
                                  Polint=np.random.uniform(0,0.7), 
                                  	#Random fractional polarization
                                  initial_angle=np.random.uniform(0,180)) 
                                  #Random angle
    I_spectra.append(Ivalues+noise_amplitude*np.random.normal(0,1,N_chan))
    Q_spectra.append(Ivalues*polarization.real+
    				noise_amplitude*np.random.normal(0,1,N_chan))
    U_spectra.append(Ivalues*polarization.imag+
    				noise_amplitude*np.random.normal(0,1,N_chan))

#Make the corresponding Stokes uncertainty arrays, for each source.
I_errors=[np.repeat(noise_amplitude,N_chan) for x in range(N_spectra)]
Q_errors=[np.repeat(noise_amplitude,N_chan) for x in range(N_spectra)]
U_errors=[np.repeat(noise_amplitude,N_chan) for x in range(N_spectra)]

#Random coordinates:
ra_array=np.random.uniform(0,360,N_spectra)
dec_array=np.random.uniform(-90,90,N_spectra)

#Create the PolSpectra table.
#Note that freq_arr is a single 1D array; the code knows it should be 2D
# or a list of arrays, so it assumes all sources have the same frequency 
#values and expands it to the appropriate dimensions.
#Setting the source_number column using range(N_spectra) gives 
#each source a unique ID number. Quantities that are common to all 
#sources, such as the flux unit and beam size, can be set for all sources
# using single values. Optional channels defined in the standard (such 
#as coordinate_system and channel_width) can also be added.
spectrumtable=polspectra.from_arrays(ra_array,dec_array,freq_arr,
                                    I_spectra,I_errors,Q_spectra,Q_errors,
                                    U_spectra,U_errors,
                                    source_number_array=range(N_spectra),
                                    beam_maj=0.01,beam_min=0.01,beam_pa=0,
                                    coordinate_system='icrs',channel_width=1e6)                                    

#Including an extra column, that isn't part of the standard:
extra_column=np.random.randint(0,100,size=N_spectra)
spectrumtable.add_column(extra_column,name='random_integer',
                         description='A random integer for each source',units='')

#Manipulate table, extracting columns and rows, and a list of column names:
print(spectrumtable[10])
print(spectrumtable['ra'])
print(spectrumtable.columns)

#Extract all data for a specific source:
print(spectrumtable[spectrumtable['source_number'] == 5])

#A verification function is provided which confirms that all columns
#have consistent numbers of channels (per row), and provides warnings
#if certain parameters (frequencies, beam sizes) seem like they may
#have the wrong units.
spectrumtable.verify_table()

spectrumtable.write_FITS('example_polspectra.fits',overwrite=True)
table_from_file=polspectra.from_FITS('example_polspectra.fits')
\end{lstlisting}

\end{document}